\documentclass[twocolumn]{aastex631}

\usepackage[T1]{fontenc}

\DeclareRobustCommand{\VAN}[3]{#2}
\let\VANthebibliography\thebibliography
\def\thebibliography{\DeclareRobustCommand{\VAN}[3]{##3}\VANthebibliography}

\newcommand{\mb}{m_{\rm b}}

\newcommand{\Mpc}{\,{\rm Mpc}}

\newcommand{\msun}{\,{\rm M_{\odot}}}

\newcommand{\Mstar}{M_*}
\newcommand{\Mdot}{{\dot M}}
\newcommand{\kms}{\,\rm km\ s^{-1}}
\newcommand{\Rvir}{R_{\rm vir}}

\newcommand{\Mhalo}{M_{\rm halo}}
\newcommand{\Mbh}{M_{\rm BH}}
\newcommand{\Mdisk}{M_{\rm disk}}

\newcommand{\MdotBH}{\Mdot_{\rm BH}}

\newcommand{\BH}{BH}
\newcommand{\BHCR}{BH+CR}
\newcommand{\CR}{NoBH+CR}
\newcommand{\NoBHCR}{NoBH}
\newcommand{\Omo}{\Omega_{\rm m}}
\newcommand{\OmL}{\Omega_{\Lambda}}
\newcommand{\Omb}{\Omega_{\rm b}}
\newcommand{\epss}{\epsilon_{\rm star}}

\shorttitle{Multi-Channel AGN Feedback in FIRE}
\shortauthors{Byrne et al.}

\begin{document}

\title{Effects of multi-channel AGN feedback in FIRE cosmological simulations of massive galaxies}

\correspondingauthor{Lindsey Byrne}
\email{byrnelin@u.northwestern.edu}

\author[0000-0002-8408-1834]{Lindsey Byrne}
\affiliation{Department of Physics and Astronomy and CIERA, Northwestern University, Evanston, IL 60201, USA}

\author[0000-0002-4900-6628]{Claude-Andr{\'e} Faucher-Gigu{\`e}re}
\affiliation{Department of Physics and Astronomy and CIERA, Northwestern University, Evanston, IL 60201, USA}

\author[0000-0002-3977-2724]{Sarah Wellons}
\affiliation{Department of Astronomy, Van Vleck Observatory, Wesleyan University, 96 Foss Hill Drive, Middletown, CT 06459, USA}

\author[0000-0003-3729-1684]{Philip F. Hopkins}
\affiliation{California Institute of Technology, TAPIR, Mailcode 350-17, Pasadena, CA 91125, USA}

\author[0000-0001-5769-4945]{Daniel Angl\'es-Alc\'azar}
\affiliation{Department of Physics, University of Connecticut, 196 Auditorium Road, U-3046, Storrs, CT 06269-3046, USA}
\affiliation{Center for Computational Astrophysics, Flatiron Institute, 162 Fifth Avenue, New York, NY 10010, USA}

\author[0000-0003-2341-1534]{Imran Sultan}
\affiliation{Department of Physics and Astronomy and CIERA, Northwestern University, Evanston, IL 60201, USA}

\author[0000-0001-6374-7185]{Nastasha Wijers}
\affiliation{Department of Physics and Astronomy and CIERA, Northwestern University, Evanston, IL 60201, USA}

\author[0000-0002-3430-3232]{Jorge Moreno}
\affiliation{Department of Physics and Astronomy, Pomona College, Claremont, CA 91711, USA}
\affiliation{Center for Computational Astrophysics, Flatiron Institute, 162 Fifth Avenue, New York, NY 10010, USA}

\author[0000-0002-7484-2695]{Sam Ponnada}
\affiliation{California Institute of Technology, TAPIR, Mailcode 350-17, Pasadena, CA 91125, USA}

% Abstract of the paper
\begin{abstract}

Feedback from supermassive black holes is believed to be a critical driver of the observed 
color bimodality of galaxies above the Milky Way mass scale. 
AGN feedback has been modeled in many galaxy formation simulations, but most implementations have involved simplified prescriptions or a coarse-grained interstellar medium (ISM). We present the first set of FIRE-3 cosmological zoom-in simulations with AGN feedback evolved to $z\sim0$, examining the impact of AGN feedback on a set of galaxies with halos in the mass range $10^{12}-10^{13} \msun$. These simulations combine detailed stellar and ISM physics with multi-channel AGN feedback including radiative feedback, mechanical outflows, and in some simulations, cosmic rays (CRs). 
We find that massive (>L*) galaxies in these simulations can match local scaling relations including the stellar mass-halo mass relation and the $\Mbh$-$\sigma$ relation; in the stronger model with CRs, they also match the size-mass relation and the Faber-Jackson relation. 
Many of the massive galaxies in the simulations with AGN feedback have quenched star formation and elliptical morphologies, in qualitative agreement with observations. In contrast, simulations at the massive end without AGN feedback produce galaxies that are too massive and form stars too rapidly, are order-of-magnitude too compact, and have velocity dispersions well above Faber-Jackson. 
Despite these successes, the AGN models analyzed do not produce uniformly realistic galaxies 
when the feedback parameters are held constant: while the stronger model produces the most realistic massive galaxies, it tends to over-quench the lower-mass galaxies.
This indicates that further refinements of the AGN modeling are needed.
\end{abstract}


%%%%%%%%%%%%%%%%%%%%%%%%%%%%%%%%%%%%%%%%%%%%%%%%%%

%%%%%%%%%%%%%%%%% BODY OF PAPER %%%%%%%%%%%%%%%%%%

\section{Introduction}
\label{sec:intro}

Over the past two decades, observations of galaxies hosting luminous quasars have found wide-angle, kiloparsec-scale, multi-phase outflows  \citep{Feruglio2010QuasarOutflows,Fischer2010iHerschel/iMarkarian231,Rupke2013BREAKINGQSO,Moe2009QUASARJ0838+2955,Liu2013ObservationsNebulae,Fabian2012ObservationalFeedback, 2017A&A...601A.143F}. These outflows provide crucial evidence for the impact of active galactic nuclei (AGN) on their host galaxies: the outflow rates can exceed the galaxy's star formation rate, and the energetics of the outflows are correlated to the luminosity of the AGN \citep{Sturm2011MASSIVE-PACS,Cicone2014MassiveObservations}.
Other properties of supermassive black holes (BHs) have also been demonstrated to correlate with properties of their host galaxies, including scaling relations between black hole mass and stellar velocity dispersion and between black hole mass and stellar bulge mass \citep{Ferrarese2000AGalaxies,Magorrian1998TheCenters,Gebhardt2000ADispersion}. These scaling relations indicate that black holes and galaxies develop together. This co-evolution can likely be explained through the feedback energy produced by accretion onto the SMBH \citep{Silk1998QuasarsFormation,DiMatteo2005EnergyGalaxies,Hopkins2007APlane}. In addition to driving winds, observational evidence suggests that AGN can heat the interstellar/circumgalactic/intracluster medium (ISM/CGM/ICM) with powerful, relativistic jets \citep{Fabian2012ObservationalFeedback}. 

 Feedback from supermassive black holes is thought to be key to understanding how galaxies quench, or stop actively forming stars \citep{DiMatteo2005EnergyGalaxies, Springel2005BlackGalaxies,Sijacki2007AFormation}. Substantial observational evidence indicates that star formation becomes increasingly likely to cease in galaxies above a stellar mass of approximately $10^{10.5} \msun$, corresponding to a halo mass $\Mhalo \sim 10^{12} \msun$, creating a population of `red and dead' massive galaxies \citep{Peng2010MASSFUNCTION,Muzzin2013THESURVEY,Tomczak2014GALAXYGALAXIES}. Understanding the physical processes that lead to this quenching is a key question in galaxy formation theory. While several possible hypotheses have been proposed, feedback from AGN has emerged as a leading candidate, with many studies showing that without AGN feedback, simulations fail to reproduce the observed galaxy mass function \citep{Somerville2008ANuclei, Harrison2017ImpactFormation} and massive galaxy properties \citep[e.g.,][]{Su2019TheGalaxies, Wellons2020MeasuringGalaxies, Parsotan2021RealisticSimulations, Cochrane2023TheSizesb}.

As the physics of SMBH accretion and feedback occurs on scales below the resolution of typical cosmological simulations, it is necessary to rely on subgrid models. However, much uncertainty remains as to how best to represent the physics involved due to open questions about the fundamental physics. Various forms of AGN feedback---including various combinations of thermal energy injection, winds, and jets---have been implemented as sub-grid models in galaxy formation simulations \citep[see reviews in][]{Somerville2015PhysicalFramework, Naab2017TheoreticalFormation, Crain2023HydrodynamicalChallenges, DiMatteo2023MassiveSimulations}. 
In addition to reproducing scaling relations, cosmological simulations with AGN feedback have demonstrated success in quenching star formation and producing quiescence in massive galaxies \citep[e.g][]{Schaye2015TheEnvironments, Dubois2016TheFeedback, Tremmel2017TheSMBHs, Dave2019Simba:Feedback, Rosito2021TheSimulations}. Such simulations have found that AGN feedback can turn star-forming disk galaxies into quenched early-type galaxies by expelling gas from the galaxy, preventing the cooling of gas in the CGM, and reducing in situ star formation \citep{Terrazas2020TheIllustrisTNG, Davies2020TheGas}.

The large-volume cosmological simulations generally utilize coarse-grained treatments of the ISM with baryonic mass resolutions in the range of $\mb \sim 10^5-10^7 \msun$, and BH models consisting of phenomenological prescriptions tuned to produce galaxy properties that match galaxy-scale observations.
Another approach is to simulate massive galaxies at much higher resolution and with more explicit small-scale physics, but to omit the cosmological setting.
Simulations of isolated massive ellipticals with AGN feedback of this type, such as those reported by e.g \cite{Ciotti2007RadiativeGas,Ciotti2010FEEDBACKFEEDBACK,Su2019TheGalaxies,Ciotti2022AOrbits, Yoon2019OnGalaxy}, have also produced promising results with regards to reproducing the properties of massive galaxies. 
An intermediate approach is to use the ``zoom-in'' technique which allows simulations to include full cosmological evolution but concentrate the resolution on halos of special interests. 
In principle, this allows the modeling of more detailed physics while retaining the cosmological context. 
Zoom-in simulations have also been used with significant success to simulate the effects of AGN feedback on the formation of massive galaxies \citep[e.g.,][]{Dubois2013AGN-drivenGalaxies, Choi2017PhysicsHoles,Choi2018TheGalaxiesc, 2020ApJ...904....8C, 2024ApJ...964...54C, Frigo2019TheGalaxies}. 
However, most previously published cosmological zoom-in simulations of massive galaxies still have relatively coarse baryonc mass resolutions $m_{\rm b} \gtrsim 10^{6}$ M$_{\odot}$ and limited sets of physical processes included.

In this paper, we build on previous work and use cosmological zoom-in simulations of galaxy formation from the Feedback In Realistic Environments (FIRE) project\footnote{See the FIRE project website at: \url{http://fire.northwestern.edu}.} to investigate the impact of multi-channel AGN feedback in relatively massive galaxies with halos in the mass range of $10^{12}-10^{13} \msun$. These correspond to Milky Way-mass ($\sim$ L*) galaxies and intermediate-mass (>L*) elliptical galaxies. In contrast to most large-volume simulations, the high resolution of the FIRE simulations allows us to resolve the multi-phase nature of the ISM and more accurately capture interactions between AGN feedback and the ISM. All simulations are evolved using the FIRE-3 code \citep{Hopkins2023FIRE-3:Simulations}. FIRE-3 represents an update from FIRE-2 \citep{Hopkins2018FIRE-2Formation}, including updates to the treatment of stellar evolution physics, nucleosynthetic yields, and low-temperature ISM cooling. Additionally, we use a detailed, multi-channel AGN feedback model which combines multiple forms of radiative feedback (e.g., radiation pressure, photo-ionization, and Compton heating) with kinetic outflows and, in some simulations, cosmic rays. 
Our highest resolution simulations at the Milky Way-mass scale have a baryonic particle mass $m_{\rm b}\approx 7,000$ M$_{\odot}$ and our highest resolution runs for >L* galaxies have $m_{\rm b}\approx3\times 10^{4}$ M$_{\odot}$. 
These resolutions are higher than even most previously published cosmological zoom-in simulations including AGN feedback at those mass scales, and the set of feedback mechanisms explicitly modeled is more complete than those used in most previous cosmological simulations.   

Our new cosmological simulations expand previous, more idealized studies of AGN feedback by our team in analytic models \citep[][]{FGQ12, FGQM12, RFG18b_analytic} and idealized simulations \citep[][]{Richings2018TheWinds, Richings21_emission} of AGN winds, as well as in simulations including the FIRE ISM physics but neglecting the cosmological environment \citep[][]{Hopkins16_concert, Torrey2020TheISM}. 
This work is complementary to \cite{Mercedes-Feliz2023LocalSimulationsc}, \cite{Mercedes-Feliz2024DenseSimulations}, and \cite{Cochrane2023TheSizesb}, which analyzed the impact of AGN winds during a luminous quasar phase in a massive FIRE-2 star-forming galaxy, and to \cite{Wellons2023ExploringSimulationsb}, which explored the impact of various models of SMBH accretion and feedback in a large suite of FIRE-2 simulations and found several physically plausible models. 
We apply AGN feedback models broadly consistent with those favored by \cite{Wellons2023ExploringSimulationsb} to a suite of cosmological zoom-in simulations with updated FIRE-3 physics. Moreover, about half of our new simulations are of order-of-magnitude higher mass resolution than the simulations of comparable-mass galaxies studied in \cite{Wellons2023ExploringSimulationsb}. 

We examine the impact of AGN feedback in the FIRE simulations by making a detailed comparison between two possible models for AGN feedback, simulations without BHs, and observations.
Our study builds on \cite{Wellons2023ExploringSimulationsb} by comparing the simulations to a broader set of observations, including more detailed structural properties.
We present a set of massive (>L*), quenched galaxies that broadly match observed scaling relations at low redshifts. Along with some of the FIRE-2 test simulations studied by \cite{Wellons2023ExploringSimulationsb}, these constitute the first such galaxies produced in cosmological simulations by the FIRE project. 
While we show examples of how including AGN feedback in simulations of massive galaxies produces much more realistic properties than simulations without BHs, we also highlight significant limitations of the AGN models analyzed in this paper. In particular, we find that while the stronger variant (with cosmic rays included) produces the most realistic >L* galaxies, it tends to over-quench lower-mass galaxies. Thus, we find evidence that further refinements of the BH growth and/or feedback physics -- perhaps including feedback parameters that depend on galaxy mass --- are warranted. 
The parameter space of possible SMBH feeding and feedback is vast, and as our results reinforce, continues to require more exploration.

The plan for this paper is as follows. In section \ref{sec:methodology} we describe the suite of zoom-in simulations analysed for this study, along with the three SMBH physics models tested. In section \ref{sec:results} we present the outcomes of the simulations with comparisons to observations of galaxy masses and star-formation rates (section \ref{sec:masses}) and structural properties (section \ref{sec:structure}). In section \ref{sec:discussion} we discuss our results, and in section \ref{sec:conclusions} we summarize our conclusions.

Throughout, we assume a standard flat $\Lambda$CDM cosmology, with matter density $\Omo \approx 0.3$, baryon density $\Omb \approx 0.05$, dark energy density $\OmL \approx 0.7$, and Hubble constant $H_0 \approx 70 \kms\Mpc^{-1}$, consistent with recent measurements \citep[][]{PlanckCollaboration2020PlanckParameters}.\footnote{In detail, the cosmological parameters in our simulations vary slightly depending on initial conditions:  $\Omo = 0.272\textup{--}0.310$, $\Omb = 0.0455$--$0.0486$, $\OmL = 0.691$--$0.728$, and $H_0=67.8$--$70.2 \kms\Mpc^{-1}$. These minor variations do not affect our conclusions.}

\section{Methods}
\label{sec:methodology}

\begin{table}
\begin{center}
\begin{tabular}{c | c  | c | c} 
Parameter & \NoBHCR & \BH & \BHCR \\ \hline \hline
v$_{\rm wind}$ [km s$^{-1}$]&  -- & 3000 &  3000 \\ \hline
$\epsilon_{\rm wind}^{\rm BH}$ & -- &   $ 5.0\times 10^{-5}$ & $ 5.0\times 10^{-5}$\\ \hline
$\epsilon_{\rm rad}^{\rm BH}$ & -- &  $0.1$ & $0.1$ \\ \hline
$\epsilon_{\rm CR}^{\rm BH}$ & -- & -- & $10^{-3}$ \\ \hline
\end{tabular}
\caption{BH feedback parameters and energetics for the physics models compared in this paper.  v$_{\rm wind}$ is the velocity of mechanical winds, $\epsilon^{\rm BH}_{\rm wind} = ( v_{\rm wind} / c)^2/2$ defines the energy going into winds (eq. \ref{eq:wind}),  
$\epsilon_{\rm rad}^{\rm BH}$ is the radiative efficiency (eq. \ref{eq:rad}), and $\epsilon_{\rm CR}$ defines the energy injected in cosmic rays (eq. \ref{eq:cr}). 
}
\label{tab:models}
\end{center}
\end{table}

\begin{table*}
\begin{center}
\begin{tabular}{c|c|c|c|c|c|c} 
Halo        &  Model        & Final $z$        & Halo Mass [$\msun$]         & Stellar Mass [$\msun$]           & $m_{\rm b}$ [$\msun$]  & $\epsilon_{\rm star}$ [pc]\\ \hline\hline
            & \NoBHCR         & 0.00     &  $1.2 \times 10^{12}$  &  $1.5 \times 10^{10}$   & $7\times10^3$          & 5   \\
\textbf{m12f}  & \BH         & 0.00     &  $1.1 \times 10^{12}$   & $ 1.5 \times 10^{10}$  & $7\times10^3$          & 10   \\
            & \BHCR        & 0.00     &  $1.0 \times 10^{12}$     & $ 8.6 \times 10^9$   & $6\times10^4$          &  10   \\ \hline
            & \NoBHCR         & 0.00     &  $1.4 \times 10^{12}$  &  $3.5 \times 10^{10}$  &  $7\times10^3$         & 5   \\
\textbf{m12q} & \BH          & 0.20     &  $1.1 \times 10^{12}$   & $ 1.4 \times 10^{10}$  & $7\times10^3$          & 5   \\
            & \BHCR        & 0.00     &  $1.2 \times 10^{12}$     & $ 7.1 \times 10^9$  & $6\times10^4$          & 10   \\ \hline
m12b        & \BH            & 0.00     &  $1.1 \times 10^{12}$   &  $2.1 \times 10^{10}$ & $7\times10^3$          & 5   \\ \hline
            & \CR         & 0.00     &      $8.8 \times 10^{11}$  &  $2.5\times 10^{10}$   & $6\times10^4$          & 10   \\
m12i        & \BH            & 0.00     &  $8.6 \times 10^{11}$   & $ 1.4 \times 10^{10}$ & $7\times10^3$          & 5   \\ 
            & \BHCR        & 0.00     &  $7.8 \times 10^{11}$    & $ 10.0 \times 10^9$   & $6\times10^4$          & 10   \\ \hline
m12m        & \BH            & 0.00     &  $1.3 \times 10^{12}$  &  $ 4.5 \times 10^{10}$ & $7\times10^3$          & 5   \\ 
            & \BHCR        & 0.00     &  $9.4 \times 10^{11}$    & $ 1.6 \times 10^{10}$  & $6\times10^4$          & 10   \\ \hline
m12r        & \NoBHCR         & 0.00     &  $8.6 \times 10^{11}$  &  $7.9\times 10^{9}$   & $7\times10^3$          & 5  \\
            & \BH            & 0.00     &  $8.9 \times 10^{11}$   & $ 8.8 \times 10^{9}$  & $7\times10^3$          & 5   \\ \hline
m12w        & \NoBHCR         & 0.00     &  $8.6 \times 10^{11}$  & $9.2\times 10^{9}$   & $7\times10^3$          & 5   \\
            & \BH            & 0.00     &  $8.9 \times 10^{11}$   & $ 1.4 \times 10^{10}$  & $7\times10^3$          & 5   \\ \hline \hline
            & \NoBHCR         & 0.00     &  $9.1 \times 10^{12}$  & $ 4.6 \times 10^{11}$    & $3\times10^5$          & 18   \\
\textbf{m13h113} & \BH       & 0.08     &  $8.1 \times 10^{12}$   & $ 1.7 \times 10^{11}$ & $3\times10^4$          & 8   \\
            & \BHCR        & 0.00     &  $6.9 \times 10^{12}$     & $ 8.9 \times 10^{10}$ &  $3\times10^5$          & 18    \\ \hline
            & \NoBHCR         & 0.00     &  $6.7 \times 10^{12}$  &  $ 3.3\times 10^{11}$   & $3\times10^5$          & 18    \\
            & \CR           & 0.50      &  $6.2 \times 10^{12}$  &   $ 5.0\times 10^{10}$   & $3\times10^5$          & 18    \\
 \textbf{m13h206} & \BH       & 0.20     &  $5.8 \times 10^{12}$   & $ 1.4 \times 10^{11}$ & $3\times10^4$          & 8   \\
            & \BHCR        & 0.00     &  $5.7\times 10^{12}$      & $ 5.2 \times 10^{10}$ & $3\times10^5$          & 18    \\ \hline
m13h002     & \BHCR        & 0.00     &  $2.0 \times 10^{13}$     &  $ 9.6 \times 10^{10}$  & $3\times10^5$          & 18    \\ \hline
m13h007     & \NoBHCR      & 0.00     &  $2.3 \times 10^{13}$     &   $ 6.4 \times 10^{11}$  & $3\times10^5$          & 18    \\
            & \BHCR        & 0.00     &  $2.0 \times 10^{13}$     & $ 1.5 \times 10^{11}$ & $3\times10^5$          & 18    \\ \hline
m13h009     & \BHCR        & 0.00     &  $2.8 \times 10^{13}$     & $ 8.0 \times 10^{10}$ & $3\times10^5$          & 18    \\ \hline
m13h029     & \NoBHCR         & 0.14     &  $1.4 \times 10^{13}$  &  $ 6.2 \times 10^{11}$   & $3\times10^5$          & 18    \\
            & \BHCR        & 0.00     &  $1.1 \times 10^{13}$     & $ 1.2 \times 10^{10}$ & $3\times10^5$          & 18    \\ \hline
m13h037     & \BHCR        & 0.00     &  $2.0 \times 10^{13}$     & $ 1.3 \times 10^{11}$ &  $3\times10^5$          & 18   \\ \hline
m13h223     & \NoBHCR         & 0.06     &  $4.1 \times 10^{13}$  &  $ 9.4 \times 10^{11}$   & $3\times10^5$          &18  \\ \hline
m13h236     & \NoBHCR         & 0.20     &  $8.9 \times 10^{12}$  &  $ 3.2 \times 10^{11}$   & $3\times10^5$          & 18    \\
            & \BHCR        & 0.00     &  $1.5 \times 10^{13}$     & $ 6.7 \times 10^{10}$ & $3\times10^5$          & 18    \\ \hline
\end{tabular}\label{tab:sims}
\caption{Parameters of the FIRE-3 simulation suite analyzed in this work. Milky Way-mass halos (`m12s') are listed in the upper half of the table and massive galaxies (`m13s') are listed in the lower half. Bolded simulations are those for which the initial conditions were run with all three primary physics models discussed in this paper. Columns are: physics model, final redshift, halo mass and stellar mass at the final redshift, the baryonic mass resolution $m_{\rm b}$, and the stellar gravitational softening length $\epsilon_{\rm star}$ in physical units.}
\end{center}
\end{table*}

\begin{figure*}
\begin{center}
	\includegraphics[width=.8\textwidth]{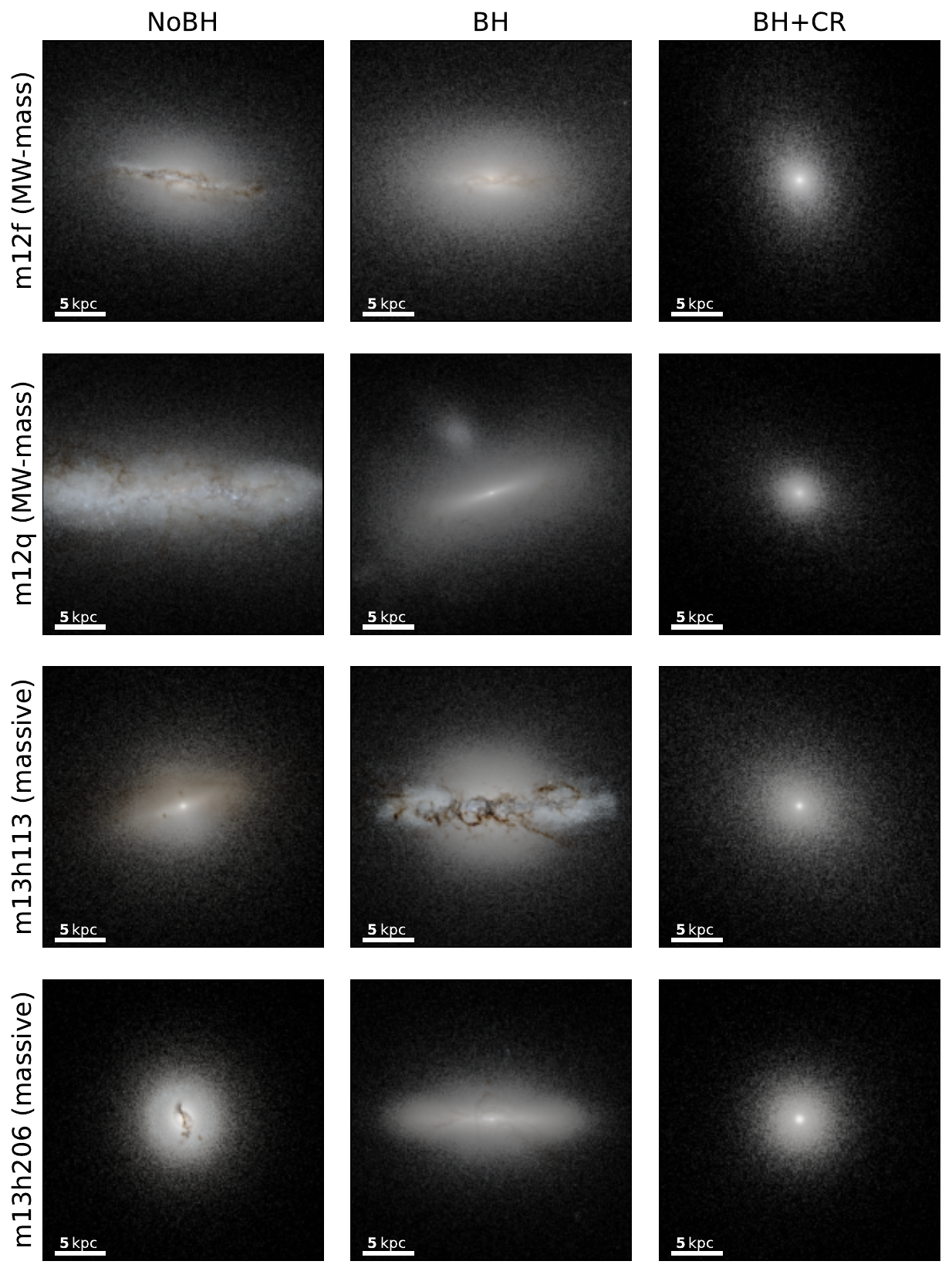}
    \caption{Edge-on mock Hubble images for two example Milky Way-mass simulated galaxies (top two rows) and two example m13 massive galaxies (bottom two rows) at $z \sim 0$. For each set of initial conditions, we compare a model with no BHs or CRs (left) to a model with BHs but no CRs (middle) and to a model with both BHs and CRs (right). Milky Way-mass galaxies tend to be disk-like at $z=0$ with the \NoBHCR~and \BH~models, while the massive galaxies produce a mix of disks and spheroidals for those two models. The \BHCR~model consistently produces elliptical galaxies.}
    \label{fig:mockHubble}
\end{center}
\end{figure*}

\begin{figure*}
	\includegraphics[width=\textwidth]{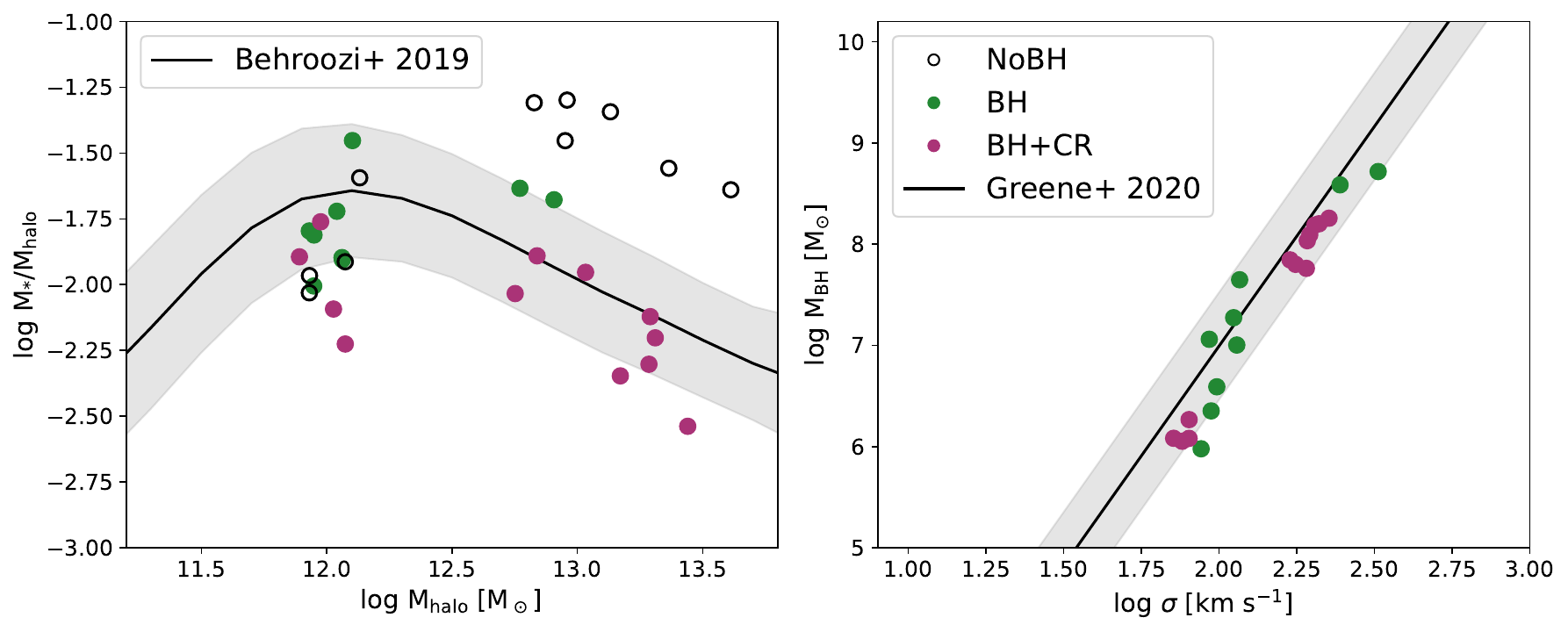}
    \caption{Stellar mass-halo mass (SMHM) relation (left) and $\Mbh-\sigma$ relation (right) for $z \sim 0$. We compare the \NoBHCR~(black, unfilled), \BH~(green), and \BHCR~(purple) models. Left panel: The SMHM relation from \cite{Behroozi2019UniverseMachine:010} and its intrinsic scatter. Right panel: The observational $\Mbh-\sigma$ relation \citep{Greene2020Intermediate-MassHoles} and its intrinsic scatter. }
    \label{fig:smhm}
\end{figure*}

\subsection{FIRE Simulations}

We analyze a set of FIRE-3 cosmological zoom-in simulations, all of which were run with the hydrodynamics code GIZMO\footnote{See the GIZMO project website at: \url{http://www.tapir.caltech.edu/~phopkins/Site/GIZMO.html}.} in meshless finite mass mode \citep{Hopkins2015AMethods}.  Details of the FIRE-3 methods and physics are explained in \cite{Hopkins2023FIRE-3:Simulations}. 
The simulations include all standard FIRE-3 physics, including multiple forms of stellar feedback, including feedback from supernovae of Type I and II, stellar winds from OB and AGB stars, photoionization, and radiation pressure. Star formation occurs in dense, self-gravitating gas. The high resolution of these simulations allows us to resolve the multiphase ISM. All simulations include magnetic fields, using the magneto-hydrodynamics methods described in \cite{Hopkins2016AccurateMagnetohydrodynamics} and \cite{Hopkins2016AMHD}. 

Our simulation suite consists of sixteen sets of initial conditions: seven Milky Way-mass galaxies with final halo mass $\Mhalo \sim10^{12} \msun$ (which we hereafter refer to as `m12' galaxies), and nine galaxies reaching $\Mhalo \sim 10^{13} \msun$ at $z=0$ (which we hereafter refer to as `m13' galaxies). Note that while we will refer to the m13 galaxies as `massive' in comparison to the Milky Way-mass galaxies, they are what would generally be considered intermediate-mass relative to >L* ellipticals. As shown in Table \ref{tab:sims}, each set of initial conditions is evolved with between one and three physics models. The three primary models we analyze for this paper are as follows: one without any supermassive black holes at all, one which models BHs with both radiative and mechanical feedback, and one with both BHs and cosmic ray physics. These models, which we hereafter refer to as the ``\NoBHCR'', ``\BH'', and ``\BHCR'' models, are described in more detail below and defined by the parameters in Table \ref{tab:models}. Four sets of initial conditions, two m12s and two m13s, were evolved with all three of these models; these galaxies are bolded in Table \ref{tab:sims}. 

We tested a range of mass resolutions, from a baryonic mass of $\mb = 7 \times 10^3 \: \msun$ for our highest-resolution simulations to a baryonic mass of $\mb = 3 \times 10^5 \: \msun$ for our lowest-resolution simulations. Simulations of higher-mass galaxies, or which include the expensive cosmic ray model, are run with lower resolution. \cite{Wellons2023ExploringSimulationsb} investigated resolution effects when examining similar FIRE simulations, also including AGN physics. 
These authors found that while there is some resolution dependence of the results, the qualitative conclusions are not affected for the range of resolutions studied here.

The final redshift is $z=0$ for most of the simulations, with a few ending at redshifts of $z \sim 0.1$ or $z \sim 0.2$; for simplicity, we refer to this range as $z \sim 0$ throughout our analysis. These minor differences in final redshift are because in some cases, evolving the simulations to $z=0$ is prohitively computationally expensive. This occurs in some massive galaxies without AGN feedback, which can develop extremely dense stellar cores when they fail to quench, and in our highest-resolution simulations of massive galaxies. While this introduces some heterogeneity to our data set, it does not affect our overall  conclusions. We note that high-resolution galaxy formation like ours are subject to substantial stochastic effects (e.g., the exact timing of bursts of star formation or AGN activity), so we could not robustly compare detailed galaxy properties between matching runs even if they were all uniformly evolved to $z=0$.

Although not the focus of this paper, we additionally considered a ``\CR'' model with cosmic rays from stellar sources but without AGN feedback (noBH+CR). We have one m12 run (m12i) and one m13 run (m13h206) with this model. However, we were unable to evolve the m13 run to $z \sim 0$, as the galaxy became unphysically dense and compact without AGN feedback. When overly-dense stellar cores develop, the simulations often become prohibitively expensive to evolve due to the short time steps required in dense regions, and this is even more so when the CR solver is enabled. 
The m13h206 NoBH+CR run became too computationally expensive to integrate past $z = 0.5$. 
Since we only have one simulation at each mass scale with the NoBH+CR model and due to stochasticity, we cannot draw robust conclusions from comparing most diagnostics on these runs to the other models. 
We therefore do not include these runs in most of our comparisons below, but we return to them in \S \ref{sec:cr} on the effects of cosmic rays. 
As we explain there, the effects on stellar sizes is one relatively robust diagnostic because once a massive simulated galaxy becomes order-of-magnitude too compact (interpreted as being due to the feedback being too weak), it is highly unlikely that the galaxy would recover realistic properties by evolving the simulation longer.

A note about the FIRE-3 simulations discussed in this study: One technical improvement in FIRE-3 (relative to FIRE-2) is the adoption of a ``velocity-aware'' technique for energy and momentum distribution to nearby gas particles in modeling supernova feedback \citep[as described in][]{Hopkins2023FIRE-3:Simulations}. 
This approach improves conservation properties when the ambient gas is not static. 
However, as Hopkins et al. (in prep.) show, this approach does not uniquely specify how much momentum should be deposited following unresolved supernova feedback events treated in subgrid. 
A particular choice was made in the implementation described in the original FIRE-3 methods paper and in the simulations in this paper. 
Although this was not highlighted in the original FIRE-3 methods paper, this implementation implies a relatively large supernova momentum injection in converging flows.\footnote{This can be understood as the energy of the pre-supernova converging flow contributing to the outward kinetic energy post-supernova.} 
In particular, converging flows result in an average supernova momentum injection that is higher in the FIRE-3 runs than in FIRE-2 runs. 
Even before AGN feedback is enabled, this leads to smaller galaxy stellar masses in massive halos compared to FIRE-2 \citep[][]{Hopkins2023FIRE-3:Simulations}. 
Since there remains significant uncertainties in the ``correct'' terminal supernova momentum to use in subgrid models and the FIRE-3 velocity-aware deposition scheme does not uniquely specify how to model it, future FIRE-3 runs will explore and use variations. Thus, some results may differ in future variants of FIRE-3 runs.

For each simulation snapshot, we define the halo center using the ``shrinking sphere'' method of \cite{Power2003TheStudy} to find the iterative center of mass of the star particles in the zoom-in region. Throughout this work, we consider the virial radius $\Rvir$ to be $R_{200c}$, the radius which encloses a sphere whose average density is 200 times the critical density. We consider the halo mass $\Mhalo$ to be $M_{200c}$, the total mass within that region.

\subsection{Black Hole Physics}
\label{sec:BHphysics}
In simulations that include BHs, the BH seeding, dynamics, accretion, and feedback follow the methods described in \cite{Hopkins2023FIRE-3:Simulations}. Black holes are probabilistically generated from gas cells and form preferentially in areas of high surface density and low metallicity. The gravitational torque-driven accretion model used in these simulations depends weakly on BH mass. The final BH masses are dominated by the accreted mass and insensitive to the initial seed mass \citep{Angles-Alcazar2013BLACKSELF-REGULATION, Angles-Alcazar2017GravitationalSimulations, Angles-Alcazar2017BlackNuclei}; for these simulations, all black hole seeds have initial mass $M_{\rm seed} =  100 \msun$. On average, one BH will seed per $M_0$ stellar mass, with $M_0 = (1-3) \times 10^6 \msun$. They move under the influence of gravity, and following the method in \cite{Wellons2023ExploringSimulationsb}, are artificially accelerated towards the local particle with the highest binding energy, which encourages them to move towards the center of the galaxy. If two black hole particles become gravitationally bound to one another within the same interaction kernel, and have overlapping force softenings, they merge within a single time-step.

Each black hole, with mass $M_{\rm BH}$ is surrounded by a subgrid accretion disk of mass $M_{\rm acc}$. 
Gas from the accretion kernel, which includes the 256 nearest gas resolution elements up to a maximum radius of 5 kpc, accretes onto the disk at a rate \begin{equation}
\dot{M}_{\rm acc} = \eta_{\rm acc} (1 - f_{w,*}) \, M_{\rm gas} \,\Omega
\end{equation}
where the efficiency parameter \begin{equation}\eta_{\rm acc} \equiv \frac{0.01[(\Mbh+M_{\rm acc})/M_{\rm disk}]^{1/6}}{1 + 3(M_{\rm gas}/\Mdisk)(\Mdisk/10^9 \msun)^{1/3}}\end{equation} is calibrated from high-resolution simulations to represent the effects of gravitational torques, which lead to angular momentum loss for gas in the galaxy's nucleus and drive gas towards the BH \citep{Hopkins2011AnHoles, Angles-Alcazar2017BlackNuclei, Angles-Alcazar2017GravitationalSimulations, Angles-Alcazar2021CosmologicalHyper-refinement}. The gas mass $M_{\rm gas}$, galaxy disk mass $\Mdisk$  
and dynamical frequency $\Omega = \sqrt{GM_{\rm enc,tot}}/R^3$ are evaluated within the radius of the accretion kernel.  $M_{\rm disk}$ is the total mass of gas and stars supported by angular momentum within the accretion kernel, and is estimated following \cite{Hopkins2011AnHoles} by decomposing the stars in the accretion kernel into isotropic and thin disk components based on their angular momentum. 
The parameter \begin{equation}f_{w,*} \equiv \left[ 1 + \frac{\bar{a}_{\rm grav}}{\langle \dot{p}_* / m_* \rangle} \right]^{-1}\end{equation} is a correction for gas lost by ejection from stellar feedback \citep{Hopkins2022WhyWhole}, where $\bar{a}_{\rm grav}$ is the gravitational acceleration inwards and $\langle \dot{p}_* / m_* \rangle \sim 10^{-7}$ cm s$^{-2}$ is the the momentum flux per unit mass from stellar feedback.

The rate at which the black hole accretes from the accretion disk, $\MdotBH$, is determined following a \cite{Shakura1973BlackAppearance.}-like $\alpha$-disk subgrid prescription:

\begin{equation}
\MdotBH \equiv \frac{M_{\rm acc}}{t_{\rm acc}}
\end{equation}
where, in our simulations,
\begin{equation}
t_{\rm acc} \equiv 42~\rm{Myr} \left[1 + \frac{\Mbh}{M_{\rm acc}} \right]^{0.4}.
\end{equation}
This effectively smooths the black hole accretion rate and allows the feedback to persist after it begins to clear the nucleus. 

Accreting BHs inject energy into their surroundings in the forms of radiative and mechanical feedback. Radiative feedback includes radiation pressure, photo-ionization, and Compton heating. Radiation transport is modelled using the same numerical methods as is used for stellar radiative feedback in the FIRE simulations \citep{Hopkins2020RadiativePhysics}.
The accretion disk emits radiation with a bolometric luminosity of 
\begin{equation}
\label{eq:rad}
\dot{E}^{\rm BH}_{\rm rad} \equiv L_{\rm bol} = \epsilon^{\rm BH}_{\rm rad} \, \dot{M}_{\rm BH} c^2 
\end{equation} 
where $\epsilon^{\rm BH}_{\rm rad} = 0.1$, and the total photon momentum flux is $\dot{p}_{\rm rad} = L_{\rm abs}/c$ where $L_{\rm abs}$ is the absorbed photon luminosity in a given gas element. The total momentum deposited by radiation in the ISM in these simulations is typically $\sim L_{\rm bol}/c$ because single scattering dominates.

\begin{figure*}
\begin{center}
	\includegraphics[width=0.8\textwidth]{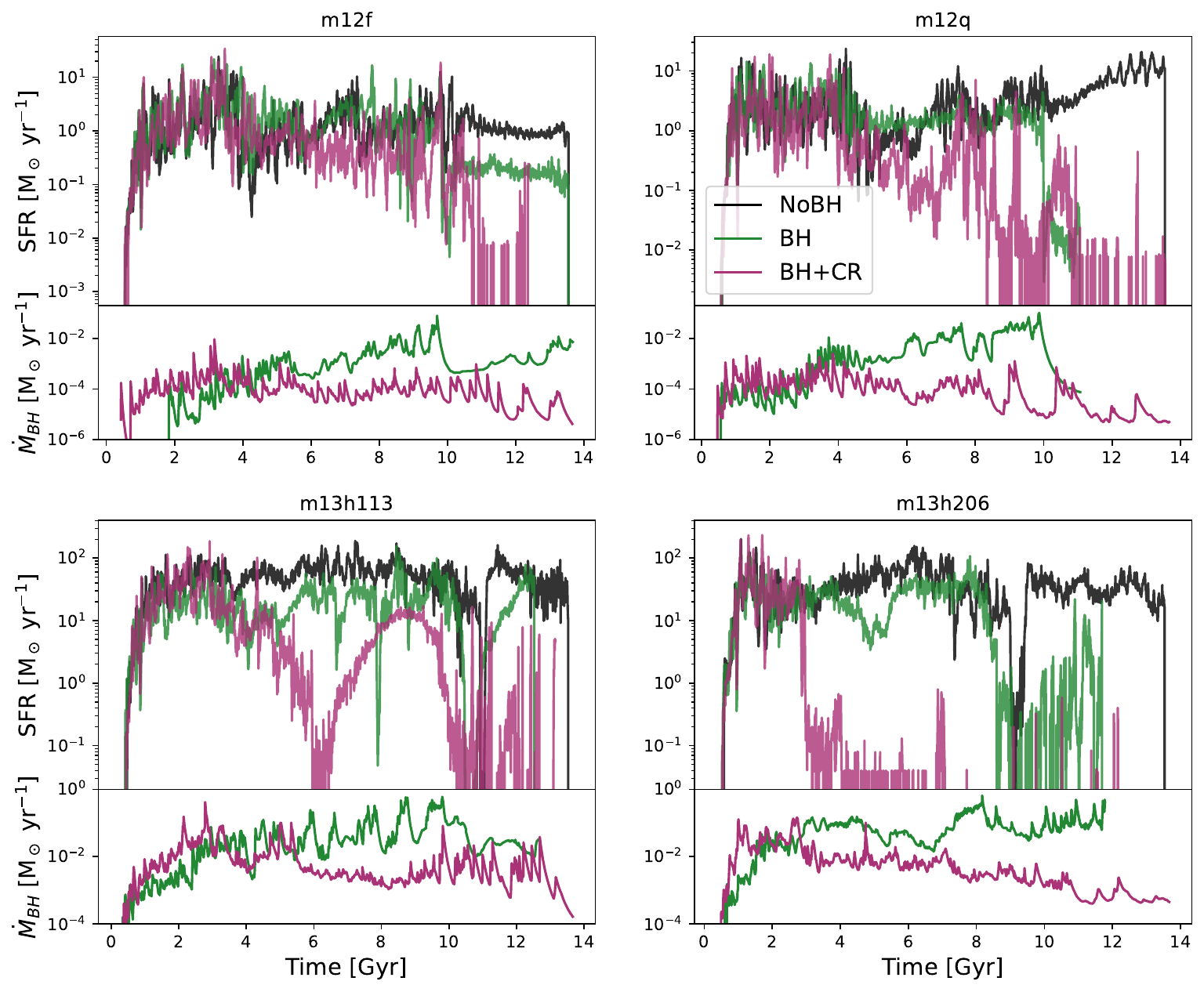}
    \caption{Star formation histories and BH accretion histories for the same four example simulated galaxies shown in Figure \ref{fig:mockHubble}: two Milky Way-mass galaxies (upper) and two massive galaxies (lower). We compare three different physics models: \NoBHCR~(black), \BH~(green), and \BHCR~(purple). The inclusion of BH feedback reduces the star formation rate at later times compared to the SFRs in simulations without BH feedback. This effect is particularly strong for the \BHCR~model in simulations of massive (m13) galaxies, in which the SFR is consistently lower than for the other models starting early in the galaxies' lifetimes. In addition to having lower SFRs than the \BH~model, the \BHCR~model also produces lower BH accretion rates.}
    \label{fig:sfr_m12f}
\end{center}
\end{figure*}

Mechanical winds are modeled by the spawning of new resolution elements, or wind cells, in the vicinity of the black hole \citep[see also][]{Richings2018TheWinds,Torrey2020TheISM,Su2021WhichGalaxiesb,Cochrane2023TheSizesb,Mercedes-Feliz2023LocalSimulationsc, Mercedes-Feliz2024DenseSimulations}. These cells, which have masses of $100 \msun$ (for m12 halos) or $400 \msun$ (for m13 halos), 
are launched with a total mass outflow rate $\dot{M}_{\rm wind} = \MdotBH$ and a velocity of $v_{\rm wind} = $ 3000 km s$^{-1}$, generating kinetic energy 
\begin{equation} \label{eq:wind} \dot{E}^{\rm BH}_{\rm wind} = \dot{M}_{\rm BH} \, v_{\rm wind}^2 / 2 = \epsilon^{\rm BH}_{\rm wind} \, \dot{M}_{\rm BH} c^2\end{equation}
where $\epsilon^{\rm BH}_{\rm wind} = ( v_{\rm wind} / c)^2/2 \approx 5 \times 10^{-5}$. The outflows are initially jet-like: collimated along the angular momentum axis of the BHs at the point of injection, though after meeting the ISM and shock heating, they tend to spread along paths of least resistance, with wide solid-angle coverage.

\subsection{Cosmic Ray Feedback}
\label{sec:CR_methods}
We also test the impact of cosmic ray feedback. The \NoBHCR~and \BH~models do not include any cosmic rays. In the ``\BHCR” model, cosmic rays are injected from both stellar and black hole sources. 

The simulations with cosmic rays use methods previously reported in detail in previous FIRE papers \citep[][]{Hopkins2021ARegimes, Hopkins2022FirstSimulationsb, Hopkins2022StandardObservations}. We summarize key elements here. We stress upfront, however, that while the CR treatment is rather complex, this is by no means a `unique’ or `final’ model of CR feedback, as the uncertainties remain large. Instead, for the purpose of this study, we use this CR model mainly as one stronger variant of the feedback model.

Most previous galaxy formation simulations that have included CRs, including previous FIRE simulations \citep[][]{Chan2019CosmicEmission, Hopkins2020ButFormation, 2021MNRAS.501.3640H} have used a `single-bin’ treatment in which CRs are represented as a second fluid with relativistic energy density in a single effective energy bin. Furthermore, most of these simulations have assumed that the CRs diffuse in space following a prescribed diffusion coefficient $\kappa$ that is constant in space and in time.

The CR runs in this paper relax both the single-bin and constant $\kappa$ assumptions. The runs explicitly evolve a CR distribution function (DF) $f({\bf x}, {\bf p}, t, s, ...)$ as a function of position {\bf x}, CR momentum {\bf p}, time $t$, and CR species $s$, using the algorithm described in \cite{Hopkins2021ARegimes}. 
 This moments-based method uses a closure akin to the M1 closure in radiation hydrodynamics and assumes that the DF is gyrotropic. The momentum-space evolution is integrated in every resolution element using the finite-momentum-space-volume method from \cite{2020MNRAS.491..993G}.

The runs in this paper evolve a CR network consisting of protons, electrons, and positrons. The injection of CRs models unresolved first-order Fermi acceleration at reverse shocks driven by mechanical feedback from SNe, stellar winds and (if applicable) AGN outflows. Specifically, CRs are injected with an initial spectrum of CR momentum $dj \propto p^{-4.2} d^3{\bf p}$, where $dj$ is the CR injection rate in an infinitesimal momentum volume element. The injection spectra are normalized such that 10\% of the initial (pre-shock) kinetic energy of the stellar feedback ejecta goes into protons and 2\% into leptons, consistent with theoretical models and empirical constraints on CR acceleration. 
For AGN feedback, if applicable, CRs are injected with an efficiency parameterized by a total energy injection rate (in the same ratio for protons and leptons)
\begin{equation} 
\label{eq:cr}
\dot{E}^{\rm BH}_{\rm CR} = \epsilon^{\rm BH}_{\rm CR} \, \dot{M}_{\rm BH} c^2,
\end{equation}
where $\epsilon^{\rm BH}_{\rm CR} = 10^{-3}$ for our runs.

The network includes numerous CR loss/gain terms, described in detail in \cite{Hopkins2022FirstSimulationsb}. These include diffusive reacceleration and adiabatic, gyro-resonant, Coulomb, ionization, hadronic, annihilation, bremsstrahlung, inverse Compton, and synchrotron loss terms.

Crucially, the CRs scatter off unresolved small-scale magnetic inhomogeneities and this drives their transport. This is what is usually modeled using a constant diffusion coefficient in galaxy formation simulations. In the simulations analyzed here, CR transport is instead predicted by evolving their DF assuming a model for the scattering coefficients, which in general depend on local plasma properties. How this is done is described in detail in \cite{Hopkins2022StandardObservations}.
Briefly, we adopt the modified external driving model described in \S 5.3.2 of \cite{Hopkins2022StandardObservations}, which is calibrated to reproduce Voyager and AMS-02 CR measurements from MeV to TeV energies in Milky Way-like simulations. This starts with an implementation of a `reference model’ which represents a theoretical understanding of how `self-confinement’ (SC) and `extrinsic turbulence’ (ET) effects developed e.g. in \cite{2013PhPl...20e5501Z}, \cite{Ruszkowski2017GLOBALSTREAMING}, \cite{zweibel:cr.feedback.review}, and \cite{2019MNRAS.485.2977T} drive CR scattering. \cite{Hopkins2022StandardObservations} show that this reference model fails to reproduce observed CR spectra and construct an additional `external driving’ term which they show produces good behavior across the diagnostics they consider. This is the transport model adopted here.

The gas and radiation in the simulations not only determine CR transport, but the CRs act on the resolved fluid. Most importantly, the CRs exert forces on the gas from the Lorentz force and from the CR pressure gradients induced by scattering.

\section{Results}
\label{sec:results}
                                                                            
\subsection{Morphologies}                                                                     

The inclusion of BH feedback affects the morphology of the simulated galaxies. Using the open-source visualization software FIREstudio \citep{Gurvich2022FIRESimulations}, we generated edge-on mock Hubble renderings of the stellar distributions of two Milky Way-mass simulated galaxy, m12q and m12f, and two massive (>L*) galaxies, m13h113 and m13h206, for each of the three models (Figure \ref{fig:mockHubble}). Milky Way-mass galaxies produce disk-like structures at $z \sim 0$ in simulations without BHs, and in simulations with BHs but no cosmic rays, while the massive galaxies produce a mix of disks and spheroidals for those two models. The addition of cosmic rays in simulations with BHs produces galaxies which appear nearly spherical regardless of mass. The presence of cosmic ray feedback in the simulated galaxies produces lower stellar masses (as discussed in section \ref{sec:masses}) and therefore lower surface brightnesses than in the galaxies without cosmic rays. As the mock images show the stellar content of the galaxies only above a minimum surface brightness, this leads the \BHCR~galaxies to appear visually smaller in these images than their counterparts with the same initial conditions, though in fact the \BHCR~galaxies are generally physically larger when considering their half-mass radii (as discussed in section \ref{sec:structure}).

\subsection{Masses and Star-Formation Rates}
\label{sec:masses}

Simulations of massive galaxies with AGN feedback can reproduce basic scaling relations at $z \sim 0$. Figure \ref{fig:smhm} shows how the simulations compare to the stellar mass-halo mass and $\Mbh-\sigma$ relations at $z \sim 0$. Points are colored by physical model: unfilled black points represent simulations without BHs (the \NoBHCR~model), green points represent simulations with BHs but without cosmic rays (the \BH~model), and purple points represent simulations with both BHs and cosmic ray physics (the \BHCR~model). 

The left panel shows the stellar mass-halo mass (SMHM) relation. Stellar masses for the simulated galaxies are calculated as all stars  within $0.1 \Rvir$ of the galaxies' centers, and are compared to the relation presented in  \cite{Behroozi2019UniverseMachine:010}. 
At $z \sim 0$, massive galaxies with no AGN feedback have become increasingly overmassive, indicating under-suppressed star formation. By contrast, galaxies with AGN feedback lie much closer to the relation. The two massive galaxies simulated with the \BH~model lie slightly (around $1 \sigma$) above the relation. Massive galaxies with the \BHCR~model lie either along the relation or slightly below it; six of the eight galaxies in this subsample are within $1 \sigma$, and the remaining two are just below $1\sigma$ beneath the relation. The inclusion of cosmic rays as another AGN feedback channel therefore appears to have a significant effect on regulating the stellar mass. 
However, a few of the Milky Way-mass galaxies with the \BHCR~model are significantly undermassive relative to observations, indicating possibly over-suppressed star formation, in contrast to the other two models which more closely reproduce the observed relation at this mass scale. As this model successfully reproduces the SMHM relation for the massive galaxies, this result may suggest that fixed AGN feedback efficiencies like the ones in our model are unable to reproduce the differential quenching at different mass scales implied by observations.

The relationship between stellar mass and stellar velocity dispersion ($\Mbh-\sigma$) is another key scaling relation, and both BH models (with and without CRs) reproduce it well. The right panel shows the $\Mbh-\sigma$ relation for the two models which included black holes, compared to the observational relation from \cite{Greene2020Intermediate-MassHoles}. The stellar velocity dispersion $\sigma$ is calculated as the standard deviation of the stellar velocities along the stellar angular momentum axis, for stars within the projected half-mass radius. Almost all of the simulations remain within the intrinsic scatter of the observed relation at $z \sim 0$; only one simulation, an m12 galaxy with the \BH~model, falls very slightly below it. 

In these simulations, BH feedback appears to regulate star formation and self-regulate BH growth. In Figure \ref{fig:sfr_m12f}, we examine the star formation histories and black hole accretion rates of the four simulated galaxies for which we can compare all three primary models. The star-formation histories are constructed using an `archaeological' method \citep[e.g. as in][]{FloresVelazquez2021TheSimulations,Gurvich2023RapidGalaxies} in which we take all stars within 0.1 $\Rvir$ at the final snapshot and estimate their formation masses by accounting for stellar mass-loss according to rates from STARBURST99 \citep{Leitherer1999Starburst99:Formation}. We then reconstruct the time-dependent SFH from the distribution of star particle formation times (SFTs) and star formation masses.  Without BH physics included, the star formation rates remain steady and relatively constant over the course of the galaxy's lifetime. The inclusion of AGN feedback leads to reductions in the star formation rates after a few Gyr of the galaxies' lifetimes: stronger feedback (i.e. the \BHCR~model) produces lower star-formation rates at late times than the \BH~model, and both produce lower rates than the \NoBHCR~model. The BH accretion rates appear to follow a similar pattern: stronger feedback in the form of our \BHCR~model leads to reduced BH accretion rates, relative to the model without cosmic rays. At times, the BH accretion rates appear to track the SF rates: remaining steady when the SF is steady, and decreasing at approximately the same time as major SF quenching episodes. While not every feature of the star formation histories appears in the BH accretion histories, this general correlation is indicative that both stars and BHs in the galactic nuclei are drawing on a common gas supply---and BH feedback reducing that gas supply can regulate both star-formation and the growth of the BH itself.

\begin{figure}
	\includegraphics[width=\columnwidth]{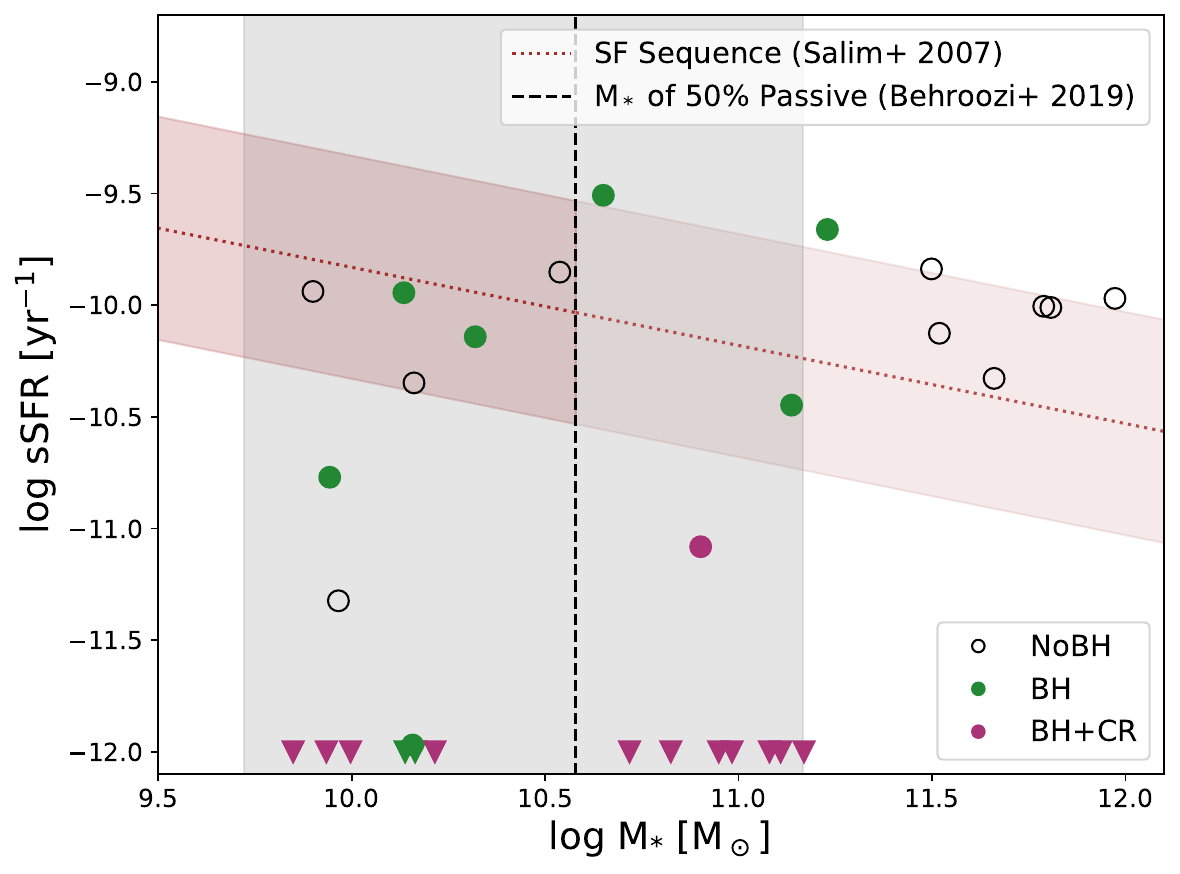}
    \caption{Specific star formation rate vs. stellar mass of the galaxy at $z \sim 0$. We compare the \NoBHCR~(black, unfilled), \BH~(green), and \BHCR~(purple) models. Triangles indicate upper limits. Observed specific star formation rates from \protect\cite{Salim2007UVUniverse} are shown for the star-forming sequence (dotted red line) along with the intrinsic observed scatter (red shaded region). We indicate the stellar mass at which 50\% of galaxies are expected to be quenched with a black dashed line \citep{Behroozi2019UniverseMachine:010}; the grey shaded region represents the inter-quartile range of the quenched fraction from 25\%-75\%. The \NoBHCR~model produces galaxies which are consistently star-forming, while the \BHCR~model produces galaxies that are consistently quenched. The \BH~model produces a mixture of quenched and star-forming Milky Way-mass galaxies; however, neither of the two most massive galaxies evolved with this model were quenched.}
    \label{fig:ssfr}
\end{figure}

\begin{figure}
	\includegraphics[width=\columnwidth]{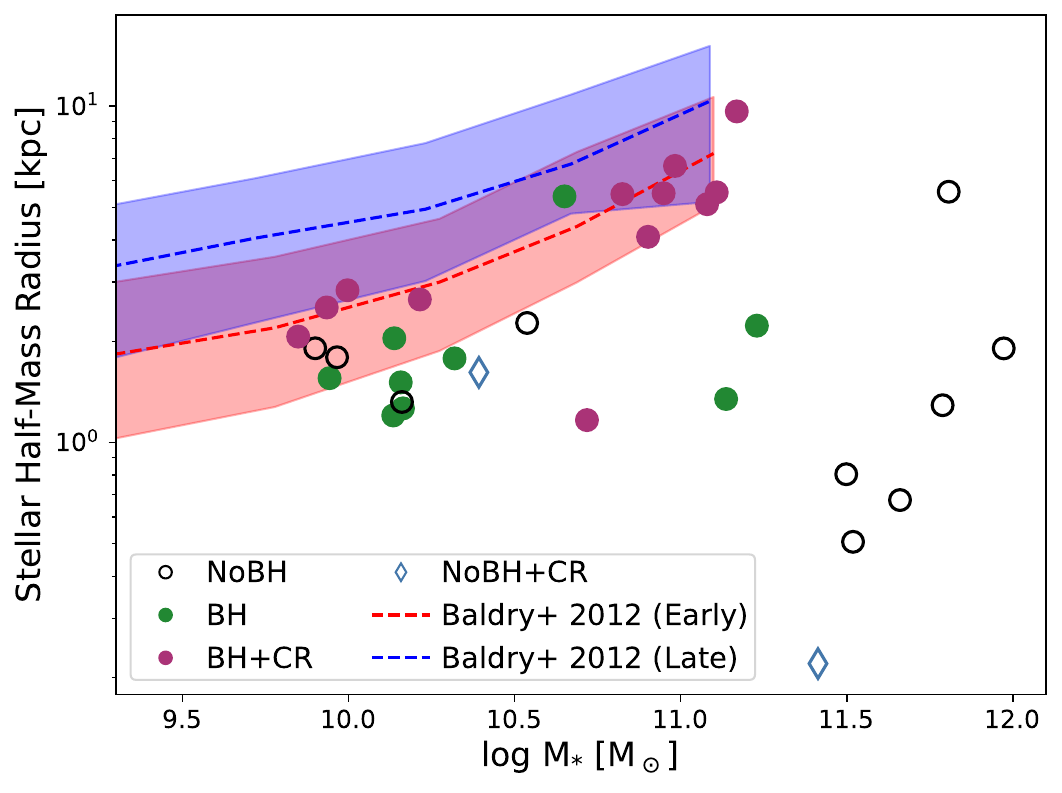}
    \caption{Projected stellar half-mass radius versus stellar mass for galaxies at redshift $z \sim 0$, compared to the observations of projected half-light radii from \protect\cite{Baldry2012Galaxy0.06}. 
    The red observational band corresponds to early-type galaxies and the blue band to late-type galaxies. 
    We compare the \NoBHCR~(black, unfilled), \BH~(green), and \BHCR~(purple) models, as well as two galaxies with the \CR~model (open blue diamonds). Massive galaxies with the \NoBHCR~model are consistently order-of-magnitude overly compact, while all but one of the simulations with the \BHCR~model are consistent with observed early-type galaxies. Massive galaxies with the \BH~model, without cosmic rays, tend to be overly compact but not to the same degree as the \NoBHCR~galaxies. The open diamond at $M_{*} \approx 10^{11.4}$ M$_{\odot}$ shows that a massive galaxy evolved to $z=0.5$ including stellar cosmic rays but no AGN feedback (the \CR~model) is order-of-magnitude too compact, similar to the \NoBHCR~model and indicating that stellar cosmic rays are not sufficient to explain stellar sizes at the massive end.
    }
    \label{fig:compactness}
\end{figure}

Figure \ref{fig:ssfr} shows the specific star formation rate for each simulation at $z \sim 0$, averaged over 10 Myr, along with observed specific star formation rates for the star-forming sequence from \cite{Salim2007UVUniverse}. We indicate the stellar mass at which 50\% of galaxies are expected to be quenched at $z=0$ \citep{Behroozi2019UniverseMachine:010}, along with the inter-quartile range of the quenched fraction from 25\%-75\%. Contrary to observations, the stronger feedback model (\BHCR) appears to quench all Milky Way-mass m12 galaxies. The \BH~and \NoBHCR~models produce a more realistic mixture of star-forming and quenched galaxies at this mass scale.  

For massive galaxies, however, observations do indicate a substantial population of quenched galaxies, and the \BHCR model is the only model to fulfill this requirement. Star formation rates for massive m13 galaxies with no AGN feedback remain at or above the star-forming sequence. Similarly, neither of the two massive galaxies produced with the \BH~model are quiescent. By contrast, all galaxies evolved with the \BHCR~ model are well below the star-forming sequences, indicating that strong feedback, such as the additional energy from cosmic rays in this model, is necessary to produce a quiescent population of massive galaxies. Many of these massive galaxies with AGN feedback and cosmic rays are quenched entirely, while a few retain small amounts of star formation, consistent with observations. 

\subsection{Structural Properties}
\label{sec:structure}

Simulations without AGN have struggled to produce realistic galaxy sizes at the massive end, as without AGN feedback, massive galaxies can become overly compact by an order of magnitude \citep[e.g.][]{Choi2017PhysicsHoles,Choi2018TheGalaxiesc,Wellons2020MeasuringGalaxies,Parsotan2021RealisticSimulations, Cochrane2023TheSizesb}. We consider the impact of AGN feedback on galaxy sizes in our simulations. Figure \ref{fig:compactness} shows the relationship between the stellar mass and the stellar half-mass radius, which is calculated as the radius at which the cumulative stellar mass profile reaches half the total stellar mass, taking the average after projecting along five randomly-selected angles. Galaxy sizes are compared to the observational $R_e-M_*$ relations found by \protect\cite{Baldry2012Galaxy0.06}, assuming a constant mass-to-light ratio, which is a reasonable approximation for old stellar populations. Simulated massive galaxies without AGN feedback (the \NoBHCR~model) are too compact by an order of magnitude; given their high stellar masses, this indicates that these galaxies are dramatically more compact than observed galaxies. By contrast, simulations with the \BHCR~model are almost all consistent with observations: all four m12s, and seven out of eight m13s, are within the scatter of the observed relation. The \BHCR~model performs better by this metric than the \BH~model, as several m12 galaxies and both m13 galaxies with the latter model are also overly compact relative to observations.

This further indicates that strong AGN feedback is necessary to produce realistic massive galaxies; without the additional energy from cosmic rays, the \BH~model we analyze cannot prevent overcooling or produce realistic galaxy sizes. To disentangle the effects of stellar vs. AGN cosmic rays, we include in Figure \ref{fig:compactness} two simulations with stellar CRs but no AGN feedback (NoBH+CR): an m12 galaxy with $\Mstar \approx 10^{10.5}$ $\msun$ and an m13 galaxy with $\Mstar \approx 10^{11.4}$ $\msun$, shown in the figure as blue diamonds. The size of the \CR~m12 galaxy is close to but slightly below the observed relation, similar to several of the \BHCR~m12 galaxies. 
The m13 massive galaxy with the \CR~model, meanwhile, is overly compact by more than an order of magnitude and not physically realistic by $z=0.5$. This run failed catastrophically and unrecoverably at this early redshift due to the computational expense of the cosmic ray model when combined with an unphysically dense stellar core. 
We conclude that the inclusion of stellar cosmic rays alone, i.e. without AGN feedback, does not produce realistic galaxies at the massive end but may produce reasonable properties at the $\sim L*$ mass scale.

\begin{figure}
	\includegraphics[width=\columnwidth]{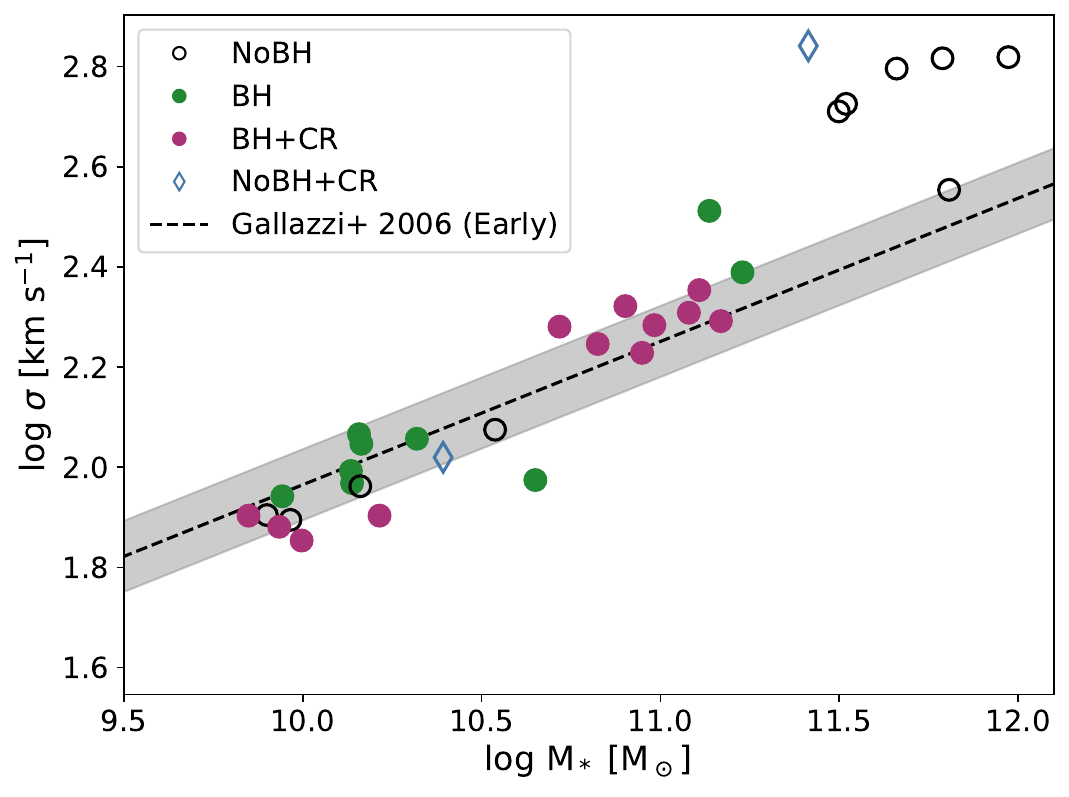}
    \caption{Stellar velocity dispersion versus stellar mass for galaxies at redshift $z \sim 0$, compared to the observed relation for early-type galaxies found by \protect\cite{Gallazzi2006AgesRelations}. We compare the \NoBHCR~(black, unfilled), \BH~(green), and \BHCR~(purple) models, as well as two galaxies with the \CR~model (open blue diamonds). 
    At the massive end, the \NoBHCR~ and \CR~simulations have clearly too high velocity dispersions, while massive galaxies with the \BHCR~model lie on or near the observed relation, and massive galaxies with the \BH~model lie slightly above it. For Milky Way-mass galaxies, all galaxies produced with the \NoBHCR~model and all but one with the \BH~model lie near the relation, while two of the four galaxies with the \BHCR~model lie slightly below it. Overall, these results show that strong AGN feedback is needed to reproduce observed velocity dispersions at the massive end, and that cosmic rays from stars alone are not sufficient.}
    \label{fig:FaberJackson}
\end{figure}

Another important scaling relation is the relationship between stellar velocity dispersion $\sigma$ and stellar mass---the Faber-Jackson relation \citep{Faber1976VelocityGalaxies.}. Figure \ref{fig:FaberJackson} shows this relation for all simulations at $z \sim 0$. While all but a few of the simulated Milky Way-mass m12 galaxies, both with and without AGN feedback, fall within the intrinsic scatter of the observed relation found by \cite{Gallazzi2006AgesRelations}, the inclusion of AGN feedback appears to have a profound impact on massive galaxies. All but one of the simulated massive galaxies without AGN feedback have overly high velocity dispersions, 0.3-0.5 dex above the intrinsic scatter of the observations. By contrast, simulated massive galaxies with the \BHCR~model lie on or near the relation, and massive galaxies with the \BH~model lay slightly (within < 0.2 dex) above it. The inclusion of BH feedback therefore appears to be key to successfully reproducing this relation for massive galaxies: the lower stellar masses and larger sizes of the galaxies with BH feedback reduce the velocity dispersions to more realistic levels, since $\sigma \varpropto \sqrt{M/R}$.  As in Figure \ref{fig:compactness}, we analyze the effects of stellar cosmic rays by including two simulations with stellar CRs but no AGN feedback, plotted as blue diamonds. The m12 galaxy with the \CR~model lies along the observed relation. The m13 \CR~galaxy has an unrealistically high stellar velocity dispersion, similar to the \NoBHCR~galaxies. The failure of the massive galaxy with the \CR~model is a further indication that AGN feedback, not stellar cosmic rays, is responsible for producing realistic galaxies at this mass scale, in agreement with Figure \ref{fig:compactness}.

\begin{figure}
	\includegraphics[width=\columnwidth]{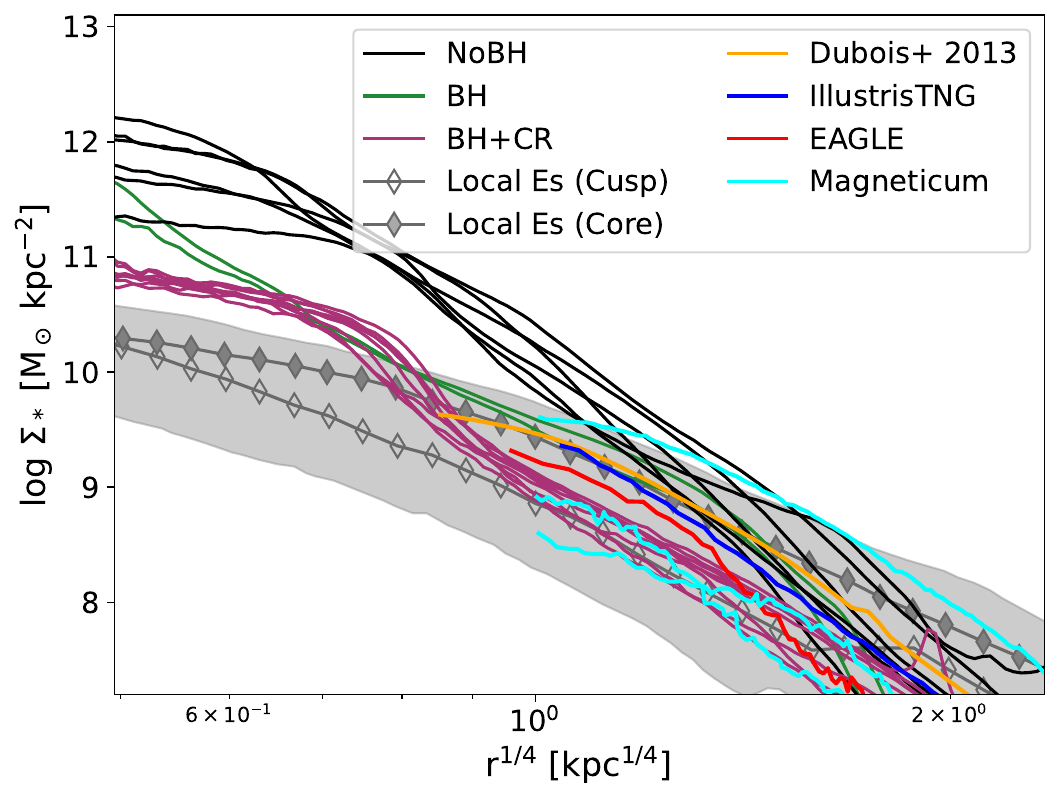}
    \caption{Stellar surface density profiles for massive galaxies with the \NoBHCR~(black, unfilled), \BH~(green), and \BHCR~(purple) models. Stellar surface densities are calculated in radial annuli and shown as a function of $r^{1/4}$. Observed profiles of local massive ellipticals (grey) are reproduced from \protect\cite{Hopkins2010ASystems}, with observations from \protect\cite{Lauer2007TheProfiles} separated into `cusp' and `core' systems. The shaded regions represent the $1\sigma$ scatter for the observed sample. The profiles of massive, early-type galaxies from other cosmological simulations with AGN feedback are also included from \cite{Dubois2013AGN-drivenGalaxies}, EAGLE  \citep{Rosito2019AssemblySimulation}, IllustrisTNG \citep{Cannarozzo2023TheIllustrisTNG}, and Magneticum \citep{Remus2022AccretedObservations}. 
    The \NoBHCR~galaxies are comparable in total stellar mass to the observed `core' galaxies (average $\Mstar = 10^{11.6} \msun$), while the \BHCR~galaxies are comparable in mass to the `cusp' galaxies (average $\Mstar = 10^{10.8} \msun$), and the \BH~galaxies lie somewhere in between.
    Simulations without BH feedback are overdense by up to two orders of magnitude in the inner regions. Introducing BH feedback reduces stellar densities: galaxies with BH feedback are overly dense in the central regions, but converge to observations by 1 kpc; with cosmic rays included in addition, the central stellar surface densities are further reduced, though still too high in the innermost regions.
    }
    \label{fig:StellarProfiles}
\end{figure}

\begin{figure*}
	\includegraphics[width=\textwidth]{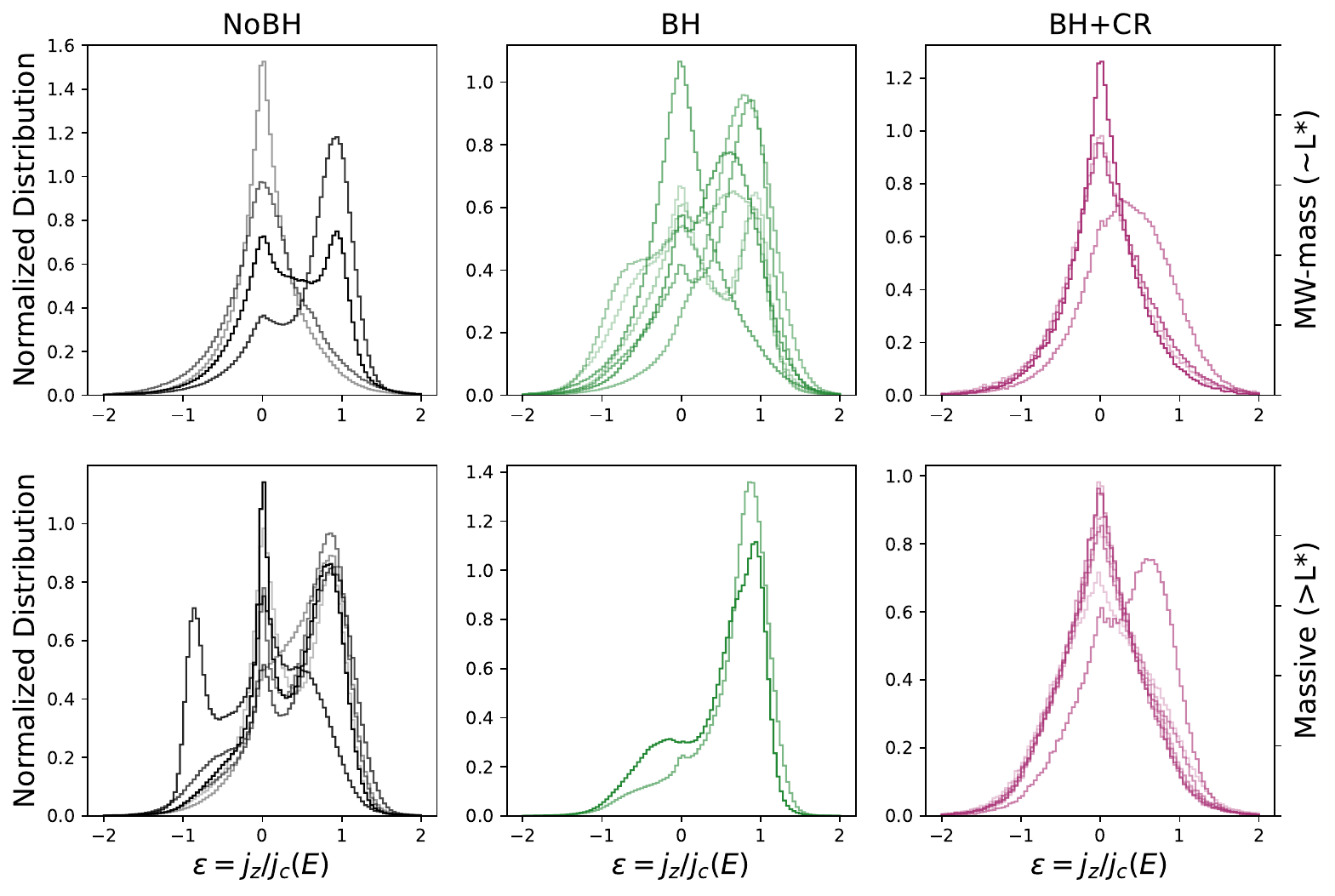}
    \caption{Mass-weighted normalized distributions of the circularity parameter $\epsilon = j_z/j_c(E)$ of star particles in each simulation. Each line represents an individual simulation. Histograms are categorized by mass scale: MW-mass m12s (top) and massive m13s (bottom), and by model: \NoBHCR~(left), \BH~(center), and \BHCR~(right). Distributions with peaks around $\epsilon = 0$ indicate elliptical morphologies, while peaks at $\epsilon = 1$ indicate disk-like morphologies.}
    \label{fig:circ}
\end{figure*}

In Figure \ref{fig:StellarProfiles}, we examine the stellar surface density profiles for the massive galaxies in our simulated sample. Projected stellar surface densities are calculated in radial annuli and shown as a function of $r^{1/4}$ \citep{deVaucouleurs1948RecherchesExtragalactiques}. 
Observed profiles of local massive ellipticals (grey) are reproduced from \cite{Hopkins2010ASystems}, where observations from \cite{Lauer2007TheProfiles} are separated into `cusp' and `core' systems 
with average stellar masses $\Mstar = 10^{10.8} \msun$ and $\Mstar = 10^{11.6} \msun$ respectively. 
Simulations with the \BHCR~model are comparable in mass to the cusp systems, while the \NoBHCR~galaxies are comparable to the core systems, and the galaxies with the \BH~model span the mass range between cusp and core.
All simulated galaxies with AGN feedback produce stellar profiles outside of 1 kpc which are in good agreement with observations. Simulations without BH feedback are overdense by up to two orders of magnitude in the inner regions, while simulations with BH feedback but no CRs are overdense by up to one order of magnitude. Introducing BH feedback with cosmic rays reduces stellar densities to more realistic levels, though they are still too high by a factor of a few in the central regions.

For context, we also compare our density profiles to profiles of massive galaxies from other cosmological simulations with AGN feedback included: the average of IllustrisTNG early-type galaxies with $10^{10.5} \msun < \Mstar < 10^{11} \msun$ \citep{Cannarozzo2023TheIllustrisTNG}, an early-type EAGLE galaxy \citep{Rosito2019AssemblySimulation}, three massive Magneticum galaxies \citep[one brightest cluster galaxy and two other early-type galaxies, selected to match in stellar mass three observed galaxies with different radial profiles: Abell 1400 BCG, NGC 3379, and NGC 2284;][]{Remus2022AccretedObservations}, and a $\Mstar = 3.2 \times 10^{11} \msun$ elliptical galaxy studied in \cite{Dubois2013AGN-drivenGalaxies}.\footnote{The surface density profile for this galaxy was inferred from the published data by integrating the three-dimensional density profile along the line-of-sight, assuming spherical symmetry.} 
Different models with AGN feedback, including ours, do comparably well at reproducing the stellar density profiles outside $\sim 1$ kpc. The high resolution of our simulations allows us to plot predicted stellar densities down to small projected radii: the stellar softening lengths for massive galaxies in our simulations are $\epss = 8$--$18$ pc, while the resolutions for the comparison simulations are $\epss=740$ pc for IllustrisTNG, $\epss=700$ pc for EAGLE \citep{Schaye2015TheEnvironments}, $\epss=700$ pc for Magneticum, and a minimum cell size $\Delta x = 500$ pc for \cite{Dubois2013AGN-drivenGalaxies}. Within a radius $r < 1$ kpc, our simulations predict significantly higher stellar densities than the observations compiled. This is a potentially significant discrepancy that our zoom-in simulations allow us to identify in a regime which could not be tested in lower-resolution simulations.

While the BHCR runs are overall the best match to the observed stellar profiles, Figure \ref{fig:StellarProfiles} suggests that these runs predict stellar `cores' with a significant change of slope at $r \sim 0.4$ kpc ($r^{1/4}\sim 0.8$ kpc$^{1/4}$). The relatively abrupt change in slope is unusual for stellar profiles, but it is on a scale substantially larger than the gravitational softening for stars in those runs ($\epsilon_{\rm star}=18$ pc). Moreover, the cores are resolved by well over $10^{4}$ stellar particles and have masses $\gtrsim 10 \times$ the masses of the central BHs. 
We also note that the NoBH runs in the figure were run with the same resolution and do not show signs of unphysical features in their stellar profiles. 
Therefore, the inner stellar profiles in the BHCR runs may be the result of complex interactions involving the baryonic processes, including the CR physics, which would be interesting to investigate more in future work.

We examine the effects of AGN feedback on the kinematic characteristics of each galaxy in Figure \ref{fig:circ}, which shows the the mass-weighted normalized distributions of the orbital circularity $\epsilon$ of every star particle within $0.1 \Rvir$ in each galaxy. Circularity $\epsilon = j_z/j_c(E)$ is defined as the ratio of each particle’s angular momentum in the $\hat{z}$ direction to that of a circular orbit with the same energy \citep{Abadi2002SimulationsDisks}. Following \cite{Scannapieco2008EffectsDiscs}, we define $j_c(E) = r v_c = r \sqrt{GM_{\rm enc}/r}$ for a particle at radius $r$, a reasonable approximation for galaxies with spherically symmetric potentials. A spheroidal galaxy will have a circularity distribution centered around $\epsilon = 0$, while for a rotation-supported disk, the distribution will peak at $\epsilon = 1$. Galaxies with both a disk and a bulge will have multiple peaks. We find that simulations without BHs, or with BHs but no CRs, tend to produce a mixture of morphologies, including many galaxies with a significant disk component. The \BHCR~ model, however, is unable to produce a realistic population of disc galaxies at $\sim L*$, providing further evidence that the feedback in this model may be too strong at that mass scale. This model consistently produces spheroidal galaxies with distributions peaking at $\epsilon = 0$, as is also visually suggested in Figure \ref{fig:mockHubble}.

\section{Discussion}
\label{sec:discussion}

\subsection{Scaling Relations and the Quenching of Star Formation}

 We demonstrate that cosmological simulations with detailed stellar and ISM physics combined with multi-channel AGN feedback can successfully produce massive galaxies with little or no star formation, broadly in agreement to observations which show galaxy populations dominated by quenched galaxies at stellar masses of $\Mstar \gtrsim 10^{10.5} \msun$. 
This is in contrast to simulations without AGN feedback which generally do not quench, especially at the massive end (the m13 galaxies in our study; see Figures \ref{fig:sfr_m12f} and \ref{fig:ssfr}). 
This provides evidence that BH feedback is necessary to produce a realistic bimodal population of galaxies. The reduction in star formation for galaxies at the massive end when AGN feedback is included is in broad agreement with many simulations using other codes \citep[e.g.][]{DiMatteo2005EnergyGalaxies,Dubois2012Self-regulatedSimulations,Schaye2015TheEnvironments,Weinberger2017SimulatingFeedback,Weinberger2018SupermassiveSimulation,Weinberger2023ActiveHaloes}. 

While it has been shown in many previous simulations that AGN feedback can produce massive galaxies with broadly realistic properties, our results are significant because none of the previously published FIRE cosmological zoom-in simulations, which include detailed stellar feedback physics, produced any quenched galaxies with realistic properties at the massive end \citep[e.g.,][]{Hopkins2014GalaxiesFormation, Hopkins2018FIRE-2Formation, Hopkins2023FIRE-3:Simulations, Wellons2020MeasuringGalaxies}. 
This was the case even when exploring the impact of various `extra physics' other than AGN feedback, including magnetic fields and cosmic rays from stellar processes \citep[e.g.,][]{Hopkins2020ButFormation}. 
The earlier FIRE study by \cite{Wellons2023ExploringSimulationsb} first analyzed a suite of FIRE-2 cosmological simulations including AGN feedback and showed preliminary successes, but the present study based on the FIRE-3 code presents a much more detailed comparison to different observational metrics. 

Beyond quenching star formation, we show that simulations of massive galaxies with AGN feedback can successfully match local scaling relations including the stellar mass-halo mass relation (Figure \ref{fig:smhm}, left), the $\Mbh$-$\sigma$ relation (Figure \ref{fig:smhm}, right), the size-mass relation (Figure \ref{fig:compactness}, and the Faber-Jackson relation ($\Mstar-\sigma$, Figure \ref{fig:FaberJackson}). Previous FIRE simulations demonstrated that in the absence of AGN feedback, massive galaxies generically become too compact and overly dense, falling well below the observed size-mass relation \citep{Wellons2020MeasuringGalaxies, Parsotan2021RealisticSimulations}. Our results (see Figures \ref{fig:compactness} and \ref{fig:StellarProfiles}) show, in agreement with some previous work, that the inclusion of AGN feedback can increase the size of massive galaxies and reduce their central stellar densities, producing much more realistic galaxy sizes \citep{Dubois2016TheFeedback,Choi2018TheGalaxiesc,Cochrane2023TheSizesb}.

As we discuss further in \S \ref{sec:mass_dep} below, our results are best viewed as a demonstration that some choices of AGN feedback parameters produce broadly realistic, quenched, elliptical galaxies at the massive end in the FIRE simulation framework. However, this is certainly not the final word on AGN feedback as we find evidence that fixed AGN parameters do not produce uniformly realistic galaxies at all mass scales -- perhaps indicating a mass dependence of AGN feedback physics.

\subsection{The Role of Cosmic Rays}
\label{sec:cr}

We find that our model of AGN feedback including cosmic rays is more efficient at quenching star formation than the AGN feedback model with only radiative and kinetic feedback channels. In the simulations presented in this paper, while both BH feedback models produce massive galaxies which better match observed properties than the model without BHs, we found that the \BHCR~model was more successful at reproducing the size-mass and $\Mstar-\sigma$ relations (Figures \ref{fig:compactness} and \ref{fig:FaberJackson}) than the \BH~model, which produced massive galaxies which were unrealistically compact and with too-high velocity dispersions. Galaxies simulated with the \BHCR~model also display more drastic reductions in their star formation rates than galaxies simulated with the \BH~model. Neither of the two massive galaxies evolved with this model became quiescent. However, the \BHCR~model in some ways appears to be too strong---it over-quenches Milky Way-mass galaxies, and consistently produces nearly spherical ellipticals across the mass range studied in this paper (Figures \ref{fig:mockHubble} and \ref{fig:ssfr}). The lack of any disk-like galaxies produced with this model, even at the Milky Way-mass scale, shows that this model has shortcomings. It is important to note that substantial uncertainty remains as to how best to model CR transport. Exploring the effect of alternate CR models, and improving this model to produce a more realistically diverse population of galaxies at different mass scales, will be critical for future work. 

Overall, our results are consistent with previous studies using FIRE simulations which found that AGN feedback in the form of CRs could have a significant impact on galaxy quenching. \cite{Wellons2023ExploringSimulationsb} found that the inclusion of CRs significantly increases the potential of a BH to suppress star formation even when the energy in CRs is relatively low. In simulations using the FIRE physics but of isolated halos (without the cosmological environment) in the mass range $\Mhalo \sim 10^{12}-10^{14}$ M$_{\odot}$, 
\cite{Su2020CosmicFail} found that CR jets suppress cooling flows more effectively than thermal heating or momentum injection, while \cite{Su2021WhichGalaxiesb} found that CR-dominated jets are able to quench galaxies with order-of-magnitude lower energetics than jets dominated by thermal or kinetic energy \citep[see also][]{2023arXiv231017692S}. It is worth noting, however, that the simulations in those works used a CR model with constant diffusion, while we use a model in which CR transport depends on local plasma properties (see \S \ref{sec:BHphysics}). In the set of simulations used in this work, CR pressure is sub-dominant in the circumgalactic medium (CGM; Sultan et al., in prep.), in contrast to constant diffusion models which can produce CGMs dominated by CRs at the L* mass scale \citep[e.g.,][]{Ji20_CR_dominated, Ji21_virial_shocks, Butsky23}.

As the \BHCR~model includes cosmic rays from both stellar and BH feedback, it is useful to determine whether these differences are driven by BH feedback or if stellar cosmic rays could be the cause. 
Fully disentangling the effects of stellar cosmic rays from BH cosmic ray feedback is beyond the scope of this work, but we comment here on what we can infer from previous work and the runs in this paper. 
In this paper, we tested a fourth physics model with cosmic rays from stellar feedback processes but without black holes (the ``\CR'' model) on one FIRE-3 m12 and one m13 galaxy (see Table \ref{tab:sims}). The m12 galaxy was integrated down to a redshift of $z=0$, while the simulation of the m13 galaxy with the \CR~model failed at $z=0.5$ after it became overly compact (see Figure \ref{fig:compactness}). While the m12 galaxy with the \CR~model produced generally realistic results, the m13 galaxy with the \CR~model showed similar properties as the m13 galaxies with the \NoBHCR~model: it failed to reproduce the observed SMHM, size-mass, and $\Mstar-\sigma$ relations due to being overmassive, too compact, and having overly high stellar velocity dispersions. It also was continuously star-forming until the final snapshot at $z=0.5$, several gigayears after all simulations with the \BHCR~model had already quenched. For these reasons, the success of the \BHCR~model at the massive end appears rely crucially on cosmic rays from BH feedback, not cosmic rays from stellar sources. 

It is worth noting that the total amounts of cosmic ray energy from stellar feedback and BH feedback are comparable in the simulations when integrated over the age of the universe. 
To see this, note that there is about one supernova per 100 M$_{\odot}$ of stars formed, so the integrated energy in CRs from stars is $E_{\rm CR}^{\star} \approx 0.1\times10^{51}~{\rm erg}~(M_{\star}/{\rm 100~M_{\odot}})=10^{59}~M_{\star,11}~{\rm erg}$, where $M_{\star,11}\equiv M_{\star} / 10^{11}~{\rm M_{\odot}}$. 
Similarly, the integrated energy in CRs from the BH is $E_{\rm CR}^{\rm BH} = \epsilon_{\rm CR}^{\rm BH} \Mbh c^{2} \approx 1.8 \times 10^{59}~{\rm erg}~M_{\rm BH,8} \epsilon_{\rm CR,-3}^{\rm BH}$, where $M_{\rm BH,8} \equiv M_{\rm BH}/ 10^{8}~{\rm M_{\odot}}$ and $\epsilon_{\rm CR,-3}^{\rm BH} \equiv \epsilon_{\rm CR}^{\rm BH}/10^{-3}$. 
This suggests that total feedback energy alone does not determine the impact on a galaxy's star formation. The lack of a CR-dominated halo for these simulations (Sultan et al., in prep.) indicates that our CR transport model produces high-energy CR outbursts which can escape from the halo, rather than the smooth CR injection into the halo seen in simulations with a constant diffusion model. These results are therefore consistent with the findings of \cite{Wellons2023ExploringSimulationsb}, who found that the responsiveness of the AGN feedback model---that is, how quickly and powerfully the BH will respond to changes in its environment, for example by creating bursts of feedback when gas in the accretion kernel is abundant---plays an important role in determining whether galaxies quench. This suggests that CRs from AGN feedback may be more efficient at suppressing star formation than CRs from stellar feedback because the AGN duty cycle produces strong outbursts of CRs, while the CR injection from stellar sources is more time-steady on average.

The successes of the \BHCR~model demonstrate that multi-channel AGN feedback can resolve some of the main failures of simulations without SMBHs in FIRE simulations with resolved ISM physics. However, we emphasize that this study analyzed only two fixed AGN feedback variants. In particular, we cannot conclude that CRs are the only way of achieving the successes we found in the \BHCR~runs. 
It is possible, for instance, that the \BHCR~model is best viewed as a ``stronger'' AGN feedback model than the specific \BH~model we analyzed. 
In a broader survey of AGN feedback parameters using FIRE-2 test simulations, \cite{Wellons2023ExploringSimulationsb} showed that while CRs from AGN feedback can efficiently quench galaxies, feedback variants in which the kinetic wind efficiency is increased can also quench massive galaxies. 
We therefore stress that further work will be needed to fully disentangle the roles of different feedback processes, including cosmic rays. 

\subsection{Evidence for a mass dependence of AGN feedback}
\label{sec:mass_dep}

In the simulations presented in this paper, the frequency with which galaxies quenched is plausibly realistic for massive galaxies but perhaps too high for Milky Way-mass galaxies (Figure \ref{fig:ssfr}). The reduced star formation in MW-mass galaxies leads to final stellar masses which are in some cases significantly below the observed SMHM relation. These results are broadly consistent with other studies of both hydrodynamic and semi-analytic models which have found that it is necessary for AGN feedback to become more efficient above a critical mass scale of $\sim L*$ for simulations to reproduce the observed bimodality \citep[e.g.][]{Croton2006TheGalaxies, Nelson2018FirstBimodality}.

Rather than enforcing an explicit mass scale for AGN activity or star-formation quenching, we chose to use fixed AGN feedback efficiencies. Our simulations capture the multiphase ISM and CGM at high resolution, and self-consistently resolve the formation of hot halos and the transition from bursty to steady star formation, both of which appear to occur at approximately the same galaxy mass as the critical mass scale for AGN feedback \citep{Bower2017TheEnd,Angles-Alcazar2017BlackNuclei,Habouzit2017BlossomsFeedback,Stern2021VirializationFeedback,Byrne2023StellarScales}. The lack of an explicitly-set mass scale in our models allows us to investigate whether the mass scales set by these processes alone could be responsible for the transition from inefficient to efficient AGN feedback. Our findings suggest that this is not the case: resolving these processes self-consistently does not appear to be sufficient to produce the critical mass scale above which AGN feedback is effective at quenching star formation, and below which it is not.

However, the subgrid treatment of BH accretion remains a source of uncertainty, with only recent cosmological hyper-refinement simulations able to explicitly model gas inflow rates down to scales $< 0.1$ pc \citep{Angles-Alcazar2021CosmologicalHyper-refinement}. In addition, our BH-feedback model is mass-invariant, in the sense that energy proportional to accretion rate is being injected independently of the Eddington ratio, the spin of the BH, the magnetic field properties, etc. We have not yet determined to what extent these types of small-scale accretion-flow physics may affect the large-scale properties of AGN feedback in our simulations. In future work, we plan to update the accretion model to incorporate insights from smaller-scale simulations about these physical processes.

\section{Summary and Conclusions}
\label{sec:conclusions}

 We present the first set of FIRE-3 simulations of massive galaxies evolved with multi-channel AGN feedback. These high-resolution, cosmological zoom-in simulations combine detailed stellar physics and a resolved multiphase ISM with multi-channel AGN feedback in the forms of radiative feedback,
 mechanical outflows, and cosmic rays. The combination of a highly-resolved ISM with detailed multi-channel stellar and AGN feedback processes distinguishes this work from existing large-volume cosmological simulations.
 
 We simulate galaxies with halos in the mass range of $10^{12}-10^{13} \msun$ down to low redshift. 
While no previous FIRE simulations including detailed stellar physics but neglecting black holes succeeded in producing realistic galaxies at the massive end of this halo mass range, we demonstrate that the addition of AGN feedback produces much improved and broadly realistic galaxy properties in several example runs of >L* galaxies evolved to $z\sim0$. 

 We test two AGN feedback models: the \BH~model, which includes both radiative and kinetic feedback channels, and the \BHCR~model, which additionally includes feedback in the form of cosmic rays. These models are both more successful at reproducing observations of massive galaxies than simulations without AGN (the \NoBHCR~model), which can fail to match observed properties such as galaxy sizes by an order of magnitude. However, the analysis reveals some limitations of the AGN feedback models: without CRs, the massive galaxies do not match all scaling relations, while the inclusion of CRs produces galaxies that are consistently quenched and spheroidal even at the lower, $\sim L$* mass scales where observations indicate that a large fraction of all galaxies are still star-forming and disky. 
 
 Our main specific results are as follows:
 
 \begin{enumerate}
    \item The addition of BH feedback affects the morphology of the galaxies. For the same set of initial conditions, a galaxy that looks like a disk in the absence of BH feedback can become spheroidal when feedback is included. Strong feedback, such as in our \BHCR~model, appears to consistently produce spheroidal galaxies, while the weaker \BH~feedback model produced a mixture of disk and spheroidal galaxies. We demonstrate this through both visual inspection of the stellar distributions and by examining the distributions of the orbital circularities of star particles. 
    \item Massive galaxies simulated with either the \BH~or the \BHCR~model can reproduce several observed low-redshift scaling relations, including the observed stellar mass-halo mass and $\Mbh-\sigma$ relations. Without AGN feedback, simulated galaxies have stellar masses which are consistently too high relative to their halo masses by a factor of 3--5.
    \item The presence of AGN feedback can quench star formation in massive galaxies. Massive galaxies in simulations without BH physics consistently remain actively star forming regardless of their mass: their star-formation rates are roughly constant throughout their lifetimes, and their specific star-formation rates remain at or above the star-forming sequence as of $z \sim 0$. By contrast, many simulations including BH feedback produce quenched galaxies. Comparing the star formation histories of galaxies with the same initial conditions evolved with each of our models, we find that the inclusion of AGN feedback leads to reductions in the star formation rates after a few Gyr of the galaxies' lifetimes: stronger feedback (i.e. the \BHCR~model) produces lower star-formation rates at late times than the \BH~model, and both produce lower rates than the \NoBHCR~model. By redshift $z \sim 0$, most massive galaxies with \BHCR~model AGN feedback are quenched.
    \item AGN feedback also produces galaxies with more realistic structural properties. We demonstrate that our stronger AGN feedback model, which includes cosmic rays, generally increases galaxy half-mass radii to be consistent with the observed size-mass relation, reduce stellar velocity dispersions to levels broadly consistent with observations of the Faber-Jackson ($\Mstar-\sigma$) relation, and reduce stellar surface densities in the central regions of the galaxies. By contrast, simulations without AGN feedback produce galaxies inconsistent with observed properties by up to an order of magnitude. The weaker \BH~model without cosmic rays is less successful at reproducing some of these relations for massive galaxies than the \BHCR~model, though it still represents a significant improvement relative to the model with no BHs at all.
\end{enumerate}
 
 There are several interesting directions for future work. 
 First, it will be necessary to continue improving our AGN feedback model in order to create a model which can produce realistic galaxies across all galaxy masses, which neither of our models currently achieve. 
 The BH accretion and feedback models used in this work are mass-invariant and our results suggest that these models do not accurately capture the differential quenching rates for galaxies at different mass scales, even with a highly-resolved ISM and self-consistently resolved hot halo formation and star formation burstiness. In future work, we plan to explore additional physics that could potentially introduce a mass dependence on the global effect of AGN feedback. For example, the Eddington ratio, BH spin, or magnetic flux could affect the accretion disk properties and therefore the intrinsic AGN feedback efficiencies, and we plan to investigate these effects. 
 Additionally, we do not yet fully understand how each separate AGN feedback channel (radiation, winds, or CRs) acts upon the galaxies and their environments, so further analysis of the effects of these individual channels will be critical. As the sample size of simulations studied in this work is relatively small, a future analysis involving a larger sample and more parameter variations would help both in disentangling the effects of different feedback mechanisms and allow for a more statistically rigorous comparisons to observations.
 
\section*{Acknowledgements}
We thank the referee for comments that helped us clarify this paper.
LB was supported by the DOE Computer Science Graduate Fellowship through grant DE-SC0020347. CAFG was supported by NSF through grants AST-2108230, AST-2307327, and CAREER award AST-1652522; by NASA through grants 17-ATP17-0067 and 21-ATP21-0036; by STScI through grant HST-GO-16730.016-A; and by CXO through grant TM2-23005X. 
Support for PFH and SP was provided by NSF Research Grants 1911233, 20009234, 2108318, NSF CAREER grant 1455342, NASA grants 80NSSC18K0562, HST-AR-15800. 
DAA acknowledges support by NSF grants AST-2009687 and AST-2108944, CXO grant TM2-23006X, JWST grants GO-01712.009-A and AR-04357.001-A, Simons Foundation Award CCA-1018464, and Cottrell Scholar Award CS-CSA-2023-028 by the Research Corporation for Science Advancement. JM is funded by the Hirsch Foundation. N.A.W. was supported by a CIERA Postdoctoral Fellowship.
Numerical calculations were run on the Caltech computer cluster Wheeler, the Northwestern computer cluster Quest, Frontera allocation FTA-Hopkins/AST20016 supported by the NSF and TACC, XSEDE allocations ACI-1548562, TGAST140023, and TG-AST140064, and NASA HEC allocations SMD-16-7561, SMD-17-1204, and SMD-16-7592.
Some figures were generated with the help of FIRE studio, an open source Python visualization package \citep{Gurvich2022FIRESimulations}. 
%%%%%%%%%%%%%%%%%%%%%%%%%%%%%%%%%%%%%%%%%%%%%%%%%%
\section*{Data Availability}

The data supporting the plots within this article are available on reasonable request to the corresponding author. 
A public version of the GIZMO code is available at \url{http://www.tapir.caltech.edu/~phopkins/Site/GIZMO.html}. Additional data including simulation snapshots, initial conditions, and derived data products are available at \url{http://fire.northwestern.edu/data/}.

\bibliographystyle{aasjournal}
\bibliography{references, references_extra} 

\end{document}